\documentclass[a4paper,
english,
11pt,
DIV11]{scrartcl}
\usepackage{microtype}%if unwanted, comment out or use option "draft"
\usepackage{textcomp}
\usepackage{latexsym} 
\usepackage{amsfonts} 
\usepackage{amssymb}
\usepackage{amsmath}
\usepackage{amsthm}
\usepackage{hyperref}
\usepackage{mathtools}%xrightharpoonup{}
\usepackage{ifthen}
\usepackage{xargs}
\usepackage{proof}
\usepackage{rta13}
\usepackage{pgf}
\usepackage{tikz}
\usepackage{wrapfig}

\theoremstyle{plain}
\newtheorem{theorem}{Theorem}
\newtheorem{lemma}[theorem]{Lemma}

\newtheorem{definition}[theorem]{Definition}
\newtheorem{example}[theorem]{Example}

\title{A Combination Framework for Complexity\footnote{This work was partially supported by FWF (Austrian Science Fund) project I-603-N18.}}

\author{Martin Avanzini${}^{1}$, and Georg Moser${}^{1}$\\
${}^{1}$ Institute of Computer Science,
University of Innsbruck,
Austria \\[-1mm]
{\small\texttt{\{martin.avanzini,georg.moser\}@uibk.ac.at}}\\
}

\begin{document}

\maketitle

\tikzstyle{dgnde}=[draw,circle,inner sep=0mm, node distance=12mm]

\begin{abstract}
In this paper we present a combination framework for
polynomial complexity analysis of term rewrite systems. The framework
covers both \emph{derivational} and \emph{runtime complexity} analysis. 
We present generalisations of powerful complexity techniques, 
notably a generalisation of \emph{complexity pairs}
and \emph{(weak) dependency pairs}. 
Finally, we also present a novel technique, called \emph{dependency graph decomposition}, 
that in the dependency pair setting greatly increases modularity.
We employ the framework in the automated complexity tool \TCT.\@
\TCT\ implements a majority of the techniques found in the literature, 
witnessing that our framework is general enough to capture a very brought setting. 
\end{abstract}

\section{Introduction}\label{s:intro}
In order to measure the complexity of a term rewrite system (TRS for short) 
it is natural to look at the maximal length of derivation sequences---the \emph{derivation length}---as
suggested by Hofbauer and Lautemann in~\cite{HL:RTA:89}. The resulting notion of complexity 
is called \emph{derivational complexity}. 
Hirokawa and the second author introduced in~\cite{HM:IJCAR:08} a variation, 
called \emph{runtime complexity}, that only takes \emph{basic} or \emph{constructor-based} terms as start terms into account. 
The restriction to basic terms allows one to accurately express the complexity of 
a program through the runtime complexity of a TRS.\@
Noteworthy both notions constitute an \emph{invariant cost model} 
for rewrite systems~\cite{AM:FLOPS:10,AM:RTA:10}.

The body of research in the field of complexity analysis of rewrite systems 
provides a wide range of different techniques to analyse the derivational and runtime complexity 
of rewrite systems, fully automatically. 
Techniques range from \emph{direct methods}, like \emph{polynomial path orders}~\cite{AM:LMCS:12,AEM:TCS:12}
and other suitable restrictions of termination orders~\cite{BCMT:JFP:2001,MSW:FSTTCS:2008}, 
to \emph{transformation techniques}, maybe most prominently adaptions of the 
\emph{dependency pair method}~\cite{HM:IJCAR:08,HM:LPAR:08,HM:IC:12,NEG:CADE:11}, \emph{semantic labeling} over 
finite carriers~\cite{A:ESSLLI:10}, methods to combine base techniques~\cite{ZK:RTA:10} 
and the \emph{weight gap principle}~\cite{HM:IJCAR:08,ZK:RTA:10}. 
Se also~\cite{Moser:2009b} for the state-of-the-art in complexity analysis of term rewrite systems.
In particular the dependency pair method for complexity analysis 
allows for a wealth of techniques
originally intended for termination analysis, foremost\emph{ (safe) reduction pairs}~\cite{HM:IJCAR:08,HM:IC:12}, 
\emph{various rule transformations}~\cite{NEG:CADE:11}, 
and \emph{usable rules}~\cite{HM:IJCAR:08,HM:IC:12} are used in complexity analysers.
Some very effective methods have been introduced specifically for 
complexity analysis in the context of dependency pairs. 
For instance, \emph{path analysis}~\cite{HM:IJCAR:08,HM:IC:12} decomposes the analysed rewrite relation
into simpler ones, by treating paths through the \emph{dependency graph} independently. 
\emph{Knowledge propagation}~\cite{NEG:CADE:11} is another complexity technique relying on 
dependency graph analysis, which allows one to propagate bounds for specific 
rules along the dependency graph. 
Besides these, various minor simplifications are implemented in tools, mostly 
relying on dependency graph analysis. 
With this paper, we provide foremost following contributions.

\begin{enumerate}
\item We propose a uniform \emph{combination framework for complexity analysis}, 
  that is capable of expressing the majority of the rewriting based complexity techniques 
  in a unified way. Such a framework is essential for the development of a modern 
  complexity analyser for term rewrite systems. 
  The implementation of our complexity analyser $\TCT$, the \emph{Tyrolean Complexity Tool}, 
  closely follows the here proposed formalisation. Noteworthy, 
  \TCT~is the only tool that is ably to participate in all 
  four complexity sub-divisions of the annual \emph{termination competition}.\footnote{C.f.~\url{http://termcomp.uibk.ac.at/}.}
\item A majority of the cited techniques were introduced in a restricted respectively
  incompatible contexts. 
  For instance, in~\cite{ZK:RTA:10} derivational complexity of relative TRSs are considered. 
  Conversely,~\cite{HM:IJCAR:08,HM:IC:12} 
  nor~\cite{NEG:CADE:11} treat relative systems, 
  and restrict their attention to \emph{basic}, aka \emph{constructor-based}, 
  start terms. 
  Where non-obvious, we generalise these techniques to our setting. 
  Noteworthy, our notion of \emph{$\PP$-monotone complexity pair} generalises 
  complexity pairs from~\cite{ZK:RTA:10} for derivational complexity, 
  \emph{$\mu$-monotone complexity pairs} for runtime complexity analysis recently 
  introduced in~\cite{HM:IC:12}, and \emph{safe reduction pairs} introduced
  in~\cite{HM:IJCAR:08} that work on dependency pairs.%
\footnote{In~\cite{NEG:CADE:11} safe reductions pairs are called \emph{$\com$-monotone reduction pairs}.}
  We also generalise the two different forms of dependency pairs for complexity analysis 
  introduced in \cite{HM:IJCAR:08} and respectively for innermost rewriting in~\cite{NEG:CADE:11}. 
  This for instance allows our tool \TCT~to employ these very powerful techniques 
  on a TRS $\RS$ relative to some theory expressed as a TRS $\SS$. 
\item Besides some simplification techniques, 
  we introduce a novel proof technique for runtime-complexity analysis, 
  called \emph{dependency graph decomposition}.
  Inspired by \emph{cycle analysis}~\cite{T:07}, this technique in principle 
  allows the analysis of two disjoint portions of the \emph{dependency graph} (almost) independently. 
  Resulting sub-problems are syntactically of simpler form, and so 
  the analysis of this sub-problems is often significantly easier. 
  Importantly, the sub-problems are usually also computational simpler 
  in the sense that their complexity is strictly smaller than the one of the input problem. 
  More precisely, if the complexity of the two generated sub-problems is bounded
  by a function in $\bigO(f)$ respectively $\bigO(g)$, then the complexity of the input 
  is bounded by $\bigO(f \cdot g)$, and often this bound is tight. 
\end{enumerate}

This paper is structured as follows. In the next 
section we cover some basics. Our combination framework 
is then introduced in Section~\ref{s:rc:framework}. 
In Section~\ref{s:rc:orders} we introduce \emph{$\mu$-monotone orders}. 
In Section~\ref{s:rc:dp} we introduce \emph{dependency pairs for complexity analysis}, 
and reprove soundness of weak dependency pairs and dependency tuples. 
In Section~\ref{s:rc:dpprocs} we introduce aforementioned simplification
and dependency graph decomposition, and finally in Section~\ref{s:rc:conclusion} 
we conclude.

\section{Preliminaries}\label{s:prel}
Let $R$ be a binary relation.
The transitive closure of $R$ is denoted by $R^+$ and
its transitive and reflexive closure by $R^{\ast}$. 
For $n\in \N$ we denote by $R^n$ the \emph{$n$-fold composition of $R$}.
The binary relation $R$ is \emph{well-founded} if 
there exists no infinite chain $a_0, a_1, \dots$ with $a_i \mathrel{R} a_{i+1}$
for all $i \in \N$. Moreover, we say that $R$ is well-founded on a set $A$ if 
there exists no such infinite chain with $a_0 \in A$.
The relation $R$ is \emph{finitely branching} if for all elements $a$, the set $\{b \mid a \mathrel{R} b\}$ is finite.
A \emph{preorder} is a reflexive and transitive binary relation. 

A (directed) \emph{hypergraph} with labels in $\mathcal{L}$ is a triple $G = (N,E,\m{lab})$ where $N$ is a set of nodes, 
$E \subseteq N \times \Pow(N)$ a set of edges, and $\ofdom{\m{lab}}{N \cup E \to \mathcal{L}}$ a \emph{labeling function}. 
For an edge $e = \tuple{u,\sseq{v}} \in E$ the node $u$ is called the \emph{source}, and the nodes $\seq{v}$ 
are called the \emph{targets} of $e$. We keep the convenient that every node is the source of at most one edge. 
If $\m{lab}(u) = l$ for node $u \in N$ then $u$ is called an \emph{$l$-node}, 
this is extended to labels $\mathcal{K} \subseteq \mathcal{L}$ in the obvious way. 
Correspondingly an edge $e \in E$ is called a \emph{$l$-edge} (respectively \emph{$\mathcal{K}$-edge})
if $\m{lab}(e) = l$ ($\m{lab}(e) \subseteq \mathcal{K})$. 
We denote by $\dtsucc[G]$ the \emph{successor relation} in $G$, 
that is, $u \dtsucc[G] v$ if there exists an edge $e = \tuple{u,\sseq{v}} \in E$
with $v \in \sseq{v}$. 
For a label $l \in \mathcal{L}$ we set $u \dtsucc[G][l] v$ (respectively $u \dtsucc[T][\mathcal{K}] v$)
 if the edge $e$ is an $l$-edge ($\mathcal{K}$-edge). 
If $u \dtsucc[G]^{*} v$ holds then $v$ is called \emph{reachable} from $u$ (in $G$). 
If every node $v \in N$ is reachable from a designated node $u$, the \emph{root} of $G$, 
then $G$ is called a \emph{hypertree}, or simply \emph{tree}. 

We assume familiarity with rewriting~\cite{BN98} and just fix notations. 
We denote by $\VS$ a countably infinite set of \emph{variables} and by $\FS$ a
\emph{signature}. 
The signature $\FS$ and variables $\VS$ are fixed throughout the paper, the 
set of terms over $\FS$ and $\VS$ is written as $\TERMS$.
Throughout the following, we suppose a partitioning of $\FS$ into \emph{constructors}
$\CS$ and \emph{defined symbols} $\DS$. The set of \emph{basic terms} $f(\seq{s})$, 
where $f \in \DS$ and arguments $s_i$ ($i= 1,\dots,n$) contain only variables or constructors, 
is denoted by $\Tb$. 
Terms are denoted by $s,t,\dots$, possibly followed by subscripts. 
We use $\subtermAt{s}{p}$ to refer to the \emph{subterm} of $s$ at position $p$. 
We denote by $\size{t}$ the \emph{size} of $t$, i.e., the number of symbols occurring in $t$.
A \emph{rewrite relation} $\to$ is a binary relation on terms closed under 
contexts and stable under substitutions.
We use $\RS,\SS,\QQ,\WW$ to refer to \emph{term rewrite systems} (\emph{TRSs} for short), i.e., 
sets of rewrite rules that obey the usual restrictions. 
We denote by $\NF(\RS)$ the \emph{normal forms} of $\RS$, and abusing notation 
we extend this notion to binary relations $\to$ on terms in the obvious way:
$\NF(\to)$ is the least set of terms $t$ such that $t \to s$ does not hold for any term $t$. 
For a set of terms $T \subseteq \TERMS$, we denote by $\fclosure{T}{\to}$ 
the least set containing $T$ and that is closed under $\to$: $s \in \fclosure{T}{\to}$ 
and $s \to t$ implies $t \in \fclosure{T}{\to}$.

For two TRSs $\QQ$ and $\RS$, 
we define $s \qrew[\QQ][\RS] t$ if there exists a context $C$, substitution $\sigma$, 
and rule $f(\seq{l}) \to r$ such that $s = C[f(l_1\sigma,\dots,l_n\sigma)]$, $t = C[r\sigma]$ 
and all arguments $l_i\sigma$ ($i = 1,\dots,n$) are $\QQ$ normal forms.
If $\QQ = \varnothing$, we sometimes drop $\QQ$ and write 
$\rew[\RS]$ instead of $\qrew[\varnothing][\RS]$. 
Note that $\rew[\RS]$ is smallest rewrite relation containing $\RS$ and thus 
corresponds to the usual definition of rewrite relation of $\RS$.
The innermost rewrite relation of a TRS $\RS$ is given by $\qrew[\RS]{\RS}$. 
We extend $\qrew[\QQ][\RS]$ to a relative setting and define
${\qrrew[\QQ]{\RS}{\SS}} \defsym {\qrss[\RS][\SS]} \cdot {\qrew[\RS][\RS]} \cdot {\qrss[\RS][\SS]}$. 

The \emph{derivation height} of a term $t$ with respect to a binary relation $\to$ on terms 
is given by $\dheight(t,\to) \max\{ n \mid \exists \seq{t}.~t \to t_1 \to \dots \to t_n\}$. 
Note that the derivation height $\dheight$ is a partial function, 
but it is defined on all inputs when 
$\to$ is well-founded and finitely branching. 
We will not presuppose that $\to$ satisfies these properties, but our 
techniques will always imply that $\dheight(t,\to)$ is well-defined. 
We emphasise that $\dheight(t,\qrrew[\QQ]{\RS}{\SS})$, if defined, 
binds the number of $\RS$ steps in all $\qrew[\QQ][\RS \cup \SS]$ 
derivations starting from $t$. 
Let $T$ be a set of terms, and define $\cc[](n,T,{\to}) \defsym \max\{ \dheight(t,\to) \mid \exists t \in T,~\size{t} \leqslant n \}$. 
The \emph{derivational complexity} of a TRS $\RS$ is given by $\dc[\RS](n) \defsym \cc[](n,\TERMS,{\rew[\RS]})$ for all $n \in \N$, 
the \emph{runtime complexity} takes only basic terms as starting terms $T$ into account: $\rc[\RS](n) \defsym \cc[](n,\Tbs,{\rew[\RS]})$ for all $n \in \N$. 
By exchanging $\rew[\RS]$ with $\qrew[\RS][\RS]$ we obtain the notions of \emph{innermost derivational} or \emph{runtime complexity} respectively.

\section{The Combination Framework}\label{s:rc:framework}

At the heart of our framework lies the notion of 
\emph{complexity processor}, or simply \emph{processor}. 
A complexity processor dictates how to transform the 
analysed input \emph{problem} into sub-problems (if any),
and how to relate the complexity of the obtained sub-problems 
to the complexity of the input problem. 
In our framework, such a processor is modeled as a set of inference rule 
$$
  \proc{\judge{\PP}{f}}{\judgeseq{\PP}{f}} \tkom
$$
over judgements of the form $\judge{\PP}{f}$. 
Here $\PP$ denotes a \emph{complexity problem} (\emph{problem} for short) 
and $\ofdom{f}{\N \to \N}$ a \emph{bounding function}. 
The validity of a judgement 
$\judge{\PP}{f}$ is given when the function $f$ binds the complexity 
of the problem $\PP$ asymptotically.

The next definition introduces complexity problems formally. 
Conceptually a complexity problem $\PP$ constitutes of a rewrite relation
$\qrrew[\QQ]{\RS}{\SS}$ represented by the \emph{finite} TRSs
$\RS,\SS$ and $\QQ$, and a set of \emph{starting terms} $\TT$. 

\begin{definition}[Complexity Problem, Complexity Function]\mbox{}\hfill
  \begin{enumerate}
  \item A \emph{complexity problem} $\PP$ (\emph{problem} for short)
  is a quadruple $\tuple{\SS,\WW,\QQ,\TT}$, in notation $\cp{\SS}{\WW}{\QQ}{\TT}$, 
  where $\SS,\WW,\QQ$ are TRSs and $\TT \subseteq \TERMS$ a set of terms.
  \item The \emph{complexity (function)} $\ofdom{\cc[\PP]}{\N \to \N}$ of $\PP$ is defined as the partial function
  $$
  \cc[\PP](n) \defsym \max\{ \dheight(t,\qrrew[\QQ]{\SS}{\WW}) \mid t \in \TT \text { and } \size{t} \leqslant n \} \tpkt
  $$ 
  \end{enumerate}
\end{definition}
Below $\PP$, possibly followed by subscripts, always denotes a complexity problem. 
Consider a problem $\PP = \cp{\SS}{\WW}{\QQ}{\TT}$.
We call $\SS$ and $\WW$ the \emph{strict} and respectively \emph{weak rules}, or \emph{component} of $\PP$. 
The set $\TT$ is called the set of \emph{starting terms} of $\PP$.
We sometimes write $l \to r \in \PP$ for $l \to r \in \SS \cup \WW$. 
We denote by $\rew[\PP]$ the rewrite relation $\qrew[\QQ][\SS \cup \WW]$. 
A derivation $t \rew[\PP] t_1 \rew[\PP] \cdots$ is also called 
a \emph{$\PP$-derivation} (\emph{starting from $t$}). 
Observe that the derivational complexity of a TRS $\RS$ 
corresponds to the complexity function of $\cp{\RS}{\varnothing}{\varnothing}{\TERMS}$. 
By exchanging the set of starting terms to basic terms 
we can express the \emph{runtime complexity} of a TRS $\RS$. 
If the starting terms are all basic terms, we call such a problem also a \emph{runtime complexity problem}.
Likewise, we can treat innermost rewriting by using $\QQ = \RS$. 
For the case $\NF(\QQ) \subseteq \NF(\SS \cup \WW)$, that is when $\rew[\PP]$ 
is included in the innermost rewrite relation of $\RS \cup \SS$, 
we also call $\PP$ an \emph{innermost complexity problem}.

\begin{example}\label{ex:pmult}
  Consider the rewrite system $\RSmult$ given by the four rules
  \begin{align*}
    \rlbl{a}\colon\ 0 + y & \to y &
    \rlbl{b}\colon\ \ms(x) + y & \to x + y & 
    \rlbl{c}\colon\ 0 \times y & \to 0 &
    \rlbl{d}\colon\ \ms(x) \times y & \to y + (x \times y) \tkom
  \end{align*}
  and let $\Tb$ denote basic terms with defined 
  symbols $+,\times$ and constructors $\ms,0$. 
  Then $\PPmult \defsym \cp{\RSmult}{\varnothing}{\RSmult}{\Tb}$
  is an innermost runtime complexity problem, in particular 
  the complexity of $\PP$ equals the innermost runtime complexity of $\RSmult$. 
\end{example}

% If the complexity function of $\PP$ is defined on all inputs and asymptotically bounded from above by 
% a linear, quadratic,\dots, polynomial function, we simply say that the complexity of $\PP$ is linear,
% quadratic,\dots, or respectively polynomial.
Note that even if $\qrrew[\QQ]{\SS}{\WW}$ is terminating, 
the complexity function is not necessarily defined on all inputs.
For a counter example, consider the problem 
$\PP_1 \defsym \cp{\SS_1}{\WW_1}{\varnothing}{\{ \m{f}(\bot) \}}$ where
$\SS_1 \defsym \{ \m{g}(\ms(x)) \to \m{g}(x) \}$ and
$\WW_1 \defsym \{ \m{f}(x) \to \m{f}(\ms(x)),\ \m{f}(x) \to \m{g}(x)\}$.
Then the set of $\qrrew[]{\SS_1}{\WW_1}$ reductions of $\m{f}(\bot)$ is
given by the family 
$$
\m{f}(\bot) \qrrew[]{\SS_1}{\WW_1} \m{g}(\ms^{n}(\bot)) \qrrsl[]{\SS_1}{\WW_1}{n} \m{g}(\bot) \tkom
$$
for $n \in \N$. So $\qrrew[]{\SS_1}{\WW_1}$ is well-founded, 
but $\dheight(\m{f}(\bot),\qrrew[]{\SS_1}{\WW_1}) = \cc[\PP_1](m)$ $(m \geqslant 2$) is undefined.
If $\qrrew[\QQ]{\SS}{\WW}$ is well-founded and \emph{finitely branching} then $\cc[\PP]$ is defined 
on all inputs, by K\"onigs Lemma.
This condition is sufficient but not necessary. 
The complexity function of the problem $\PP_2 \defsym \cp{\SS_2}{\WW_1}{\varnothing}{\{ \m{f}(\bot) \}}$, 
where $\SS_2 \defsym \{\m{g}(x) \to x \}$, is constant
but $\m{f}(\bot) \qrrew[]{\SS_2}{\WW_1} \ms^n(\bot)$ for all $n \in \N$, 
i.e, $\qrrew[]{\SS_2}{\WW_1}$ is not finitely branching.

In this work we do not presuppose that the complexity function is defined on all inputs, 
instead, this will be determined by our methods. 
To compare partial functions in proofs, we use \emph{Kleene equality}:
two partial functions $\ofdom{f,g}{\N \to \N}$ are equal, in notation $f \keq g$, if for all $n \in \N$
either $f(n)$ and $g(n)$ are defined and $f(n) = g(n)$, or both $f(n)$ and $g(n)$ are undefined. 
We abuse notation and write $f \kgeq g$ if for all $n \in \N$ with $f(n)$ defined, 
$f(n) \kgeq g(n)$ with $g(n)$ defined holds.
Then $f \keq g$ if and only if $f \kgeq g$ and $g \kgeq f$. 

\begin{definition}[Judgement, Processor, Proof]\mbox{}\hfill
  \begin{enumerate}
  \item A\emph{ (complexity) judgment} is a statement $\judge{\PP}{f}$ where $\PP$ is a complexity problem 
    and $\ofdom{f}{\N \to \N}$.
    The judgment is \emph{valid} if $\cc[\PP]$ is defined on all inputs, and $\cc[\PP] \in {\bigO(f)}$. 
  \item 
    A \emph{complexity processor} $\PROC$ (\emph{processor} for short) is an inference rule 
    $$
    \proc[\PROC]{\judge{\PP}{f}}{\judgeseq{\PP}{f}} \tkom
    $$
    over complexity judgements. 
    The problems $\seq{\PP}$ are called the \emph{sub-problems generated by $\PROC$ on $\PP$}. 
    The processor $\PROC$ is \emph{sound} 
    if 
    $$
    \cc[\PP_1] \in \bigO(f_1) \land \cdots \land \cc[\PP_n] \in \bigO(f_n) \IMPLIES \cc[\PP] \in \bigO(f) \tkom
    $$
    it is \emph{complete} 
    if 
    % $$
    % \cc[\PP_1] \in \bigO(f_1) \land \cdots \land \cc[\PP_n] \in \bigO(f_n) \FOLLOWSFROM \cc[\PP] \in \bigO(f) \tpkt
    % $$
    the inverse direction holds.
  \item 
    Let $\mathsf{empty}$ denotes the axiom $\judge{\cp{\varnothing}{\WW}{\QQ}{\TT}}{f}$ for all TRSs $\WW$ and $\QQ$, set of terms $\TT$
    and $\ofdom{f}{\N \to \N}$. 
    A \emph{complexity proof} (\emph{proof} for short) of a judgement $\judge{\PP}{f}$ 
    is a deduction from the axiom $\mathsf{empty}$ and \emph{assumptions} 
    $\judge{\PP_1}{f_1}, \dots, \judge{\PP_n}{f_n}$ using sound processors, 
    in notation $\judge[\jdgmt{\PP_1}{f_1}, \dots, \jdgmt{\PP_n}{f_n}]{\PP}{f}$.
  \end{enumerate}
\end{definition}

We say that a complexity proof is \emph{closed} if its set of assumptions is empty, otherwise it is \emph{open}. 
We follow the usual convention and annotate side conditions as premises to inference rules. 
Soundness of a processor guarantees that validity of judgements is preserved.
Completeness ensures that a deduction can give asymptotically tight bounds.
The next lemma verifies that our formal system is correct. 

\begin{lemma}
  If there exists a closed complexity proof $\judge{\PP}{f}$, then the judgement $\judge{\PP}{f}$ is valid.
\end{lemma}
\begin{proof}
  The lemma follows by a standard induction on the length of proofs, exploiting 
  that the underlying set of processors is sound. 
\end{proof}

\section{Orders for Complexity}\label{s:rc:orders}

Maybe the most obvious tools for complexity analysis in rewriting
are \emph{reduction orders}, in particular \emph{interpretations}.
Consequently these have been used quite early for complexity analysis. 
For instance, in~\cite{BCMT:JFP:2001} \emph{polynomial interpretations}
are used in a direct setting in order to estimate the runtime complexity analysis of a TRS.\@ 
In~\cite{ZK:RTA:10} so called \emph{complexity pairs} are employed to estimate
the derivational complexity in a relative setting. 
Weakening monotonicity requirements of complexity pairs gives rise to a 
notion of \emph{reduction pair}, so called \emph{safe reduction pairs}~\cite{HM:IJCAR:08}, 
that can be used to estimate the runtime complexity of dependency pair problems, 
cf.\ also~\cite{HM:IC:12,NEG:CADE:11}.
In the following, we introduce \emph{$\PP$-monotone complexity pairs}, that give
a unified account of the orders given in~\cite{BCMT:JFP:2001,ZK:RTA:10,HM:IC:12,NEG:CADE:11}. 
% As in \emph{$\mu$-monotone interpretation} from~\cite{HM:IC:12}, 
% the parameter $\mu$ is used to control monotonicity of the underlying orders. 

% \subsection{$\PP$-Monotone Complexity Pairs}
We fix a complexity problem $\PP = \cp{\SS}{\WW}{\QQ}{\TT}$. 
Consider a proper order $\succ$ on terms, and let $\G$ denote 
a mapping associating a term with a natural number.
Then $\succ$ is \emph{$\G$-collapsible} (on $\PP$)
if $\G(s) > \G(t)$ whenever $s \qrrew[\QQ]{\SS}{\WW} t$ and $s \succ t$ holds 
for all terms $s \in \fclosure{\TT}{\rew[\PP]}$ reachable from $\TT$.
If in addition 
$\G(t)$ is asymptotically bounded by a function $\ofdom{f}{\N \to \N}$ 
in the size of $t$ for all start terms $t \in \TT$, 
i.e., $\G(t) \in \bigO(f(\size{t}))$ for $t\in\TT$, 
we say that $\succ$ \emph{induces} the complexity 
$f$ on $\PP$. 
We note that only in pathological cases an order is not collapsible on $\PP$. 
In particular polynomial and matrix interpretations~\cite{BN98,HW:RTA:06} are collapsible, 
and also \emph{recursive path order}~\cite{BN98} are. 
All these termination techniques have also been suitable tamed 
so that the induced complexity can be given by a polynomial~\cite{BCMT:JFP:2001,MMNWZ:CAI:11,AEM:TCS:12}, 
for runtime and partly also for derivational complexity. 

Consider an order $\succ$ that induces the complexity $f$ on $\PP$.
If this order includes the relation $\qrrew[\QQ]{\SS}{\WW}$, 
the judgement $\judge{\PP}{f}$ is valid. 
To check the inclusion, we consider
\emph{complexity pair} $(\succsim, \succ)$, 
where $\succsim$ denotes a preorder $\succsim$ 
compatible with $\succ$, in the sense that ${\succsim} \cdot {\succ} \cdot {\succsim} \subseteq {\succ}$
holds. 
It is obvious that when both orders are monotone and stable under substitutions, 
the assertions $\WW \subseteq {\succsim}$ and $\SS \subseteq {\succ}$ imply
${\qrrew[\QQ]{\SS}{\WW}} \subseteq {\succ}$ as desired. 
Monotonicity ensures that the orders are closed under contexts. 
Of course, closure under context is only required on argument positions 
that can be rewritten in reductions of staring terms $\TT$. 
Such argument positions are called \emph{usable} below. 
In~\cite{HM:IC:12} it is shown how to precisely capture reductions from
basic terms by \emph{context sensitive rewriting}~\cite{L:SOFSEM:95}, 
and conclusively \emph{$\mu$-monotone orders}~\cite{Z:RTA:97}, can be exploited 
for runtime complexity analysis. 

The parameter $\mu$ denote a \emph{replacement map}, i.e.,\ a map that
assigns to every $n$-ary function symbol $f \in \FS$ 
a subset of its argument positions: $\mu(f) \subseteq \set{1,\dots,n}$.
In the realm of context sensitive rewriting, the replacement map 
governs under which argument positions a rewrite step is allowed.
Here we use $\mu$ to designate which argument positions are usable. 
Denote by $\Pos[\mu](t)$ the \emph{$\mu$-replacing positions} in $t$, defined as
$\Pos[\mu](t) \defsym \set{\posempty}$ if $t$ is a variable,
and $\Pos[\mu](t) \defsym \set{\posempty} \cup \set{\text{$i \posc p \mid i \in \mu(f)$ and $p \in \Pos[\mu](t_i)$}}$ 
for $t = f(\seq{t})$.
% \begin{equation*}
%   \Pos[\mu](t) \defsym
%   \begin{cases}
%     \set{\posempty} & \text {if $t$ is a variable,} \\
%     \set{\posempty} \cup \set{\text{$i \posc p \mid i \in \mu(f)$ and $p \in \Pos[\mu](t_i)$}} & \text{if $t = f(\seq{t})$.}
%   \end{cases}
% \end{equation*}
For a binary relation $\to$ on terms we denote by $\muTA{\mu}{\to}$ 
the least set of terms $s$ such that whenever $\subtermAt{s}{p} \not\in\NF(\to)$, 
then $p$ is a $\mu$-replacing position.
The following constitutes an adaption of \emph{usable replacement maps}~\cite{HM:IC:12} to our setting. 
\begin{definition}
  Let $\PP$ be a complexity problem with starting terms $\TT$ and
  let $\RS$ denote a set of rewrite rules. 
  A replacement map $\mu$ is called a \emph{usable replacement map} for $\RS$ in $\PP$, if 
  $\fclosure{\TT}{\rew[\PP]} \subseteq \muTA{\mu}{\qrew[\QQ][\RS]}$.
\end{definition}
Hence a usable replacement map $\mu$ of $\RS$ in $\PP$ determines under which argument 
positions an $\RS$ step is possible. 
% If $i \in \mu(f)$ for all replacement map $\mu$ of $\RS$ in $\PP$, we also say 
% the argument position $i$ of $f$ is a usable position of $\RS$ in $\PP$. 
This is illustrated by the following example. 
\begin{example}[Example~\ref{ex:pmult} continued]\label{ex:pmult:usable}
  Consider the derivation
  \begin{multline*}
  \underline{\ms(0) \times \ms(\ms(0))} 
  \rew[\PPmult] \ms(\ms(0)) + (\underline{0 \times \ms(\ms(0))})
  \rew[\PPmult] \underline{\ms(\ms(0)) + 0}
  \rew[\PPmult] \ms(\underline{\ms(0) + 0}) \rew[\PPmult] \cdots
  \tkom
  \end{multline*}
  where redexes are underlined.
  Observe that e.g.\ multiplication occurs only under the second argument position of addition in 
  this sequence. 
  The replacement map $\mu_\times$, defined by $\mu_\times(+) = \{ 2 \}$ 
  and $\mu_\times(\times) = \mu_\times(\ms) = \varnothing$, thus constitutes 
  a usable replacement map for the multiplication rules $\{\rlbl{c},\rlbl{d}\}$ in $\PPmult$.
  Since the argument position of $\ms$ is not usable in $\mu_\times$, the last step witnesses 
  that $\mu_\times$ does not designate a usable replacement map for the addition rules $\{\rlbl{a},\rlbl{b}\}$. 
\end{example}

We say that an order $\succ$ is \emph{$\mu$-monotone} if it is monotone on $\mu$ positions, 
in the sense that for all function symbols $f$, 
if $i \in \mu(f)$ and $s_i \succ t_i$ then 
$f(s_1, \dots, s_i, \dots, s_n) \succ f(s_1, \dots, t_i, \dots, s_n)$ 
holds. 
Consider a usable replacement map $\mu$ for a TRS $\RS$ in $\PP$, 
and let $\succ$ be a $\mu$-monotone order stable under substitution. 
Compatibility of $\succ$ with $\RS$ ensures that every $\RS$ 
step in a $\PP$ derivation of $t \in \TT$ is oriented by $\succ$:
\begin{lemma}\label{l:rc:monotonecp}
  Let $\mu$ be a usable replacement map for $\RS$ in $\PP$, and let
  $\succ$ denote a $\mu$-monotone order that is stable under substitutions. 
  If $\RS \subseteq {\succ}$ holds, i.e., rewrite rules in $\RS$ are oriented from 
  left to right, then $s \qrew[\QQ][\RS] t$ implies $s \succ t$ for all 
  terms $s \in \fclosure{\TT}{\rew[\PP]}$. 
\end{lemma}
\begin{proof}
      Since $\mu$ is a usable replacement map for $\RS$ in $\PP$, 
  it suffices to show the claim for $s \in \muTA{\mu}{\qrew[\QQ][\RS]}$. 
  Suppose $s \qrew[\QQ][\RS,p] t$, hence $p \in \Pos[\mu](t)$. 
  We show that for every prefix $q$ of $p$, $\subtermAt{s}{q} \succ \subtermAt{t}{q}$ holds.
  Consider a prefix $q$ of $p$. The proof is by induction on $\size{p} - \size{q}$.
  The base case $q = p$ is covered by compatibility and stability under substitutions.
  For the inductive step, consider 
  prefix $q \posc i$ of $p$, 
  where by induction hypothesis $\subtermAt{s}{q \posc i} \succ \subtermAt{t}{q \posc i}$. 
  Since $p \in \Pos[\mu](s)$ and $q \posc i$ is a prefix of $p$ 
  it is not difficult to see that $i \in \mu(f)$. 
  Thus 
  $$
  \subtermAt{s}{q} = f(s_1,\dots,\subtermAt{s}{q \posc i},\dots,s_n)
  \succ f(s_1,\dots,\subtermAt{t}{q \posc i},\dots,s_n) = \subtermAt{s}{q} \tkom
  $$
  follows by $\mu$-monotonicity of $\succ$. 
  From the claim, the lemma is obtained using $q = \posempty$. 
\end{proof}

The next definition of 
\emph{$\PP$-monotone complexity pair} $(\succsim,\succ)$
ensures that $\succsim$ and $\succ$ are sufficiently monotone, in 
the sense that the inclusions $\WW \subseteq {\succsim}$ 
on weak and $\SS \subseteq {\succ}$ on strict rules of $\PP$ 
extend to $\qrew[\QQ][\WW]$ respectively $\qrew[\QQ][\SS]$, 
according Lemma~\ref{l:rc:monotonecp}.

\begin{definition}[Complexity Pair, $\PP$-monotone]\mbox{}\hfill
  \begin{enumerate}
  \item A \emph{complexity pair} is a pair $(\succsim,\succ)$, such that 
    $\succsim$ is a stable preorder and $\succ$ a stable order with 
    ${\succsim} \cdot {\succ} \cdot {\succsim} \subseteq {\succ}$. 
  \item 
    Suppose $\succsim$ is $\mu_\WW$-monotone for a usable replacement map of $\WW$ in $\PP$, 
    and likewise $\succ$ is $\mu_\SS$-monotone for a usable replacement map of $\SS$ in $\PP$. 
    Then $(\succsim,\succ)$ is called \emph{$\PP$-monotone}.
  \end{enumerate}
  % The pair $(\succsim,\succ)$ is \emph{compatible} with $\PP$ if 
  % $\succsim$ is compatible with weak rules $\WW$, and $\succ$ compatible with strict rules:
  % $\SS \subseteq {\succ}$ and $\WW \subseteq {\succsim}$. 
\end{definition}
When the set of starting terms is unrestricted, as for derivational complexity analysis, 
then trivially only the full replacement map is usable for rules of $\PP$. 
In this case, our notion of complexity pair collapses to the one given by Zankl and Korp~\cite{ZK:RTA:10}.
We emphasise that in contrast to~\cite{HM:IC:12}, our notion of 
complexity pair is parameterised in separate replacements 
for $\succsim$ and $\succ$. By this separation we can recast 
\emph{(safe) reduction pairs} originally proposed in~\cite{HM:IJCAR:08}, 
employed in the dependency pair setting below, as instances of complexity pairs (cf. Lemma~\ref{l:rc:mucom}). 

It is undecidable to determine if $\mu$ is a usable replacement map for rules $\RR$ in $\PP$.
Exploiting that for runtime-complexity the set of starting terms consists of basic
terms only, Hirokawa and Moser~\cite{HM:IC:12} define an algorithm that computes 
good approximations of usable replacement maps $\mu$, both for the case of full 
and innermost rewriting.
The next lemma verifies that complexity pairs bind the length of derivations. 
\begin{lemma}
  Consider a $\PP$-monotone complexity pair $(\succsim,\succ)$ that is $\G$-collapsible 
  on $\PP$, 
  and compatible in the sense that $\WW \subseteq {\succsim}$ and $\SS \subseteq {\succ}$.
  Then $\dheight(t,\qrrew[\QQ]{\SS}{\WW})$ is defined for all $t \in \TT$, 
  in particular $\dheight(t,\qrrew[\QQ]{\SS}{\WW}) \leqslant \G(t)$.
\end{lemma}
\begin{proof}
  Consider an arbitrary derivation 
  $t = t_0 \qrrew[\QQ]{\SS}{\WW} t_1 \qrrew[\QQ]{\SS}{\WW} t_2 \qrrew[\QQ]{\SS}{\WW} \cdots$
  for some term $t \in \TT$. 
  Consider a sub-sequence $t_i \qrrew[\QQ]{\SS}{\WW} t_{i+1}$ for some $i \in \N$. 
  Using compatibility and $\PP$-monotonicity we see 
  $t_i \succsim^* \cdot \succ \cdot \succsim^* t_{i+1}$ by
  Lemma~\ref{l:rc:monotonecp}, 
  using that $\succsim$ is a preorder we even have 
  $t_i \succsim \cdot \succ \cdot \succsim t_{i+1}$ 
  and hence $t_i \succ t_{i+1}$ by definition of complexity pair.
  Since $\succ$ is $\G$-collapsible on $\qrrew[\QQ]{\SS}{\WW}$ we conclude 
  $\G(t) = \G(t_0) > \G(t_1) > \G(t_2) > \cdots$. 
  Conclusively the sequence $t_0,t_1,t_2,\dots$ is finite, and its length is bounded by $\G(t)$. 
  The lemma follows. 
  % $$
  % t_i = s_{i,1} \qrew[\QQ][\WW] \cdots \qrew[\QQ][\WW] s_{i,n} \qrew[\QQ][\SS] s_{i,n+1} \qrew[\QQ][\WW] \cdots \qrew[\QQ][\WW] s_{i,n+m} = t_{i+1}\tkom
  % $$
  % for $i \in \N$. 
  % By assumption $s_{i,k} \in \muTA{\mu}{\PP}$ for $k = 1,\dots,n+m$. 
  % Then by 
  % By compatibility, and exploiting that $\succsim$ and $\succ$ are $\mu$-monotone 
  % and stable under substitutions, a standard induction gives that 
  % $$
  % t_i = s_{i,1} \succsim \cdots \succsim s_{i,n} \succ s_{i,n+1} \succsim \cdots \succsim s_{i,n+m} = t_{i+1}\tkom
  % $$
  % holds. Since $\succsim$ is a preorder, we thus have $t_i \succsim \cdot \succ \cdot \succsim t_{i+1}$
  % and so $t_i \succ t_{i+1}$ holds for all $i \in \N$. 
  %
  % Since $\succ$ is $\G$-collapsible on $\qrrew[\QQ]{\SS}{\WW}$ we conclude 
  % $\G(t) = \G(t_0) \succ \G(t_1) \succ \G(t_2) \cdots$. 
  % Conclusively the sequence $t_0,t_1,t_2,\dots$ is finite, and its length is bounded by $\G(t)$. 
  % The lemma follows. 
\end{proof}

As an immediate consequence we obtain our first processor. 
\begin{theorem}[Complexity Pair Processor]\label{rc:proc:rp}
  Let $(\succsim,\succ)$ be a $\PP$-monotone complexity pair 
  such that $\succ$ induces the complexity $f$ on $\PP$. 
  The following processor is sound:
  $$
  \proc[CP]
  {\judge{\cp{\SS}{\WW}{\QQ}{\TT}}{f}}
  {\SS \subseteq {\succ} & \WW \subseteq {\succsim}} \tpkt
  $$
\end{theorem}

A variation of the complexity pair processor, that 
iterative orients disjoint subsets of $\SS$, occurred first in~\cite{ZK:RTA:10}. 
The following processor constitutes a straight forward generalisation
of~\cite[Theorem~4.4]{ZK:RTA:10} to our setting.
\begin{theorem}[Decompose Processor~\cite{ZK:RTA:10}]\label{rc:proc:decompose}
  The following processor is sound:
  $$
  \proc[decompose]
    {\judge{\cp{\SS_1 \cup \SS_2}{\WW}{\QQ}{\TT}}{f + g}}
    {\judge{\cp{\SS_1}{\SS_2 \cup \WW}{\QQ}{\TT}}{f} 
      & \judge{\cp{\SS_2}{\SS_1 \cup \WW}{\QQ}{\TT}}{g}
    } \tpkt
  $$
\end{theorem}
\begin{proof}[Proof Outline]
  Obviously the number of $\SS_1 \cup \SS_2$ steps in a derivation 
  is given by the sum of $\SS_1$ and respectively $\SS_2$ steps in this derivation. 
\end{proof}

The effect of applying the complexity pair processor on one of the resulting sub-problems, 
for instance $\cp{\SS_1}{\SS_2 \cup \WW}{\QQ}{\TT}$, is to \emph{move}
the strictly oriented rules $\SS_1$ to the weak component.
This is demonstrated in the following proof, that was automatically found by our complexity prover \TCT.\@
The following proof was found by our tool \TCT\ demonstrates this.

\begin{example}[Examples~\ref{ex:pmult} and~\ref{ex:pmult:usable} continued]
  Consider the linear polynomial interpretation $\AS$ over $\N$ such that 
  $0_\AS= 0$, $\ms_\AS(x) = x$, $x +_\AS y = y$ and $x \times_\AS y = 1$. 
  Let $\PP_{\m{c}} \defsym \cp{\{\rlbl{c}\}}{\{\rlbl{a},\rlbl{b},\rlbl{d}\}}{\RSmult}{\Tb}$ 
  denote the problem that accounts for the rules $\rlbl{c}\colon\ {0 \times y} \to y$ in $\PPmult$. 
  The induced order $>_\AS$ together with its reflexive closure $\geqslant_\AS$
  forms a $\PP_{\m{c}}$-monotone complexity pair $({\geqslant_\AS},{>_\AS})$
  that induces linear complexity on $\PP_{\m{c}}$.
  Here monotonicity can be shown using the replacement maps given in Example~\ref{ex:pmult:usable}.
  The following depicts a complexity proof
  $\judge[\jdgmt{\cp{\{\rlbl{a},\rlbl{b},\rlbl{d}\}}{\{\rlbl{c}\}}{\RSmult}{\Tb}}{g}]{\PPmult}{\lambda n.n + g}$. 
  $$
  \proc[decompose]
    {\judge{\PPmult}{\lambda n.n + g}}
    { \proc[CP]
      {\judge{\cp{\{\rlbl{c}\}}{\{\rlbl{a},\rlbl{b},\rlbl{d}\}}{\RSmult}{\Tb}}{\lambda n.n} }
      {\{\rlbl{c}\} \subseteq {>_\AS} & \{\rlbl{a},\rlbl{b},\rlbl{d}\} \subseteq {\geqslant_\AS}}
      & 
      \judge{\cp{\{\rlbl{a},\rlbl{b},\rlbl{d}\}}{\{\rlbl{c}\}}{\RSmult}{\Tb}}{g} 
    } \tpkt
  $$
\end{example}
The above complexity proof can now be completed iterative, on the simpler 
problem $\cp{\{\rlbl{a},\rlbl{b},\rlbl{d}\}}{\{\rlbl{c}\}}{\RSmult}{\Tb}$. 
Since the complexity of $\PPmult$ is bounded below by a quadratic polynomial, 
one however has to use a technique beyond linear polynomial interpretations. 

We remark that the decompose processor finds applications beyond its combination with complexity pairs, 
our tool \TCT\ uses this processor for instance for the separation of independent components in 
the dependency graph.

\section{Dependency Pairs for Complexity Analysis}\label{s:rc:dp}
The introduction of the \emph{dependency pair} (\emph{DP} for short)~\cite{AG:TCS:00}, 
and its prominent reincarnation in the \emph{dependency pair framework}~\cite{T:07}, 
drastically increased power and modularity in termination provers.
It is well established that the DP method is unsuitable for complexity analysis. 
The induced complexity is simply too high~\cite{AS:12},
in the sense that the complexity of $\RS$ is not suitably reflected in its canonical DP problem. 
Hirokawa and Moser~\cite{HM:IJCAR:08} recover this deficiency with 
the introduction of \emph{weak dependency pairs}.
Crucially, weak dependency pairs group different function calls 
in right-hand sides, using \emph{compound symbols}.

In this section, we first introduce a notion of \emph{dependency pair complexity problem} (\emph{DP problem} for short), 
a specific instance of complexity problem. 
In Theorem~\ref{rc:proc:wdp} and Theorem~\ref{rc:proc:tuples} we
then introduce the \emph{weak dependency pair} and \emph{dependency tuples} processors, 
that construct from a runtime-complexity problem its canonical 
DP problem. 
We emphasise that both processors are conceptually \emph{not} new, 
weak dependency pairs were introduced in~\cite{HM:IJCAR:08}, and the seminal paper of Noschinski~et~al.~\cite{NEG:CADE:11} 
introduced dependency tuples. 
In this work we provide alternative proofs of 
central theorems from~\cite{HM:IJCAR:08} and~\cite{NEG:CADE:11}. 
Unlike in the cited literature, we show a direct simulation
that also accounts for weak rules, conclusively our processors provide a 
generalisation of~\cite{HM:IJCAR:08,NEG:CADE:11}.

Consider a signature $\FS$ that is partitioned into defined symbols $\DS$ 
and constructors $\CS$. 
Let $t \in \TERMS$ be a term. 
For $t = f(\seq{t})$ and $f \in \DS$, we set $\mrk{t} = \mrk{f}(\seq{t})$ where $\mrk{f}$
is a new $n$-ary function symbol called \emph{dependency pair symbol}. 
For $t$ not of this shape, we set $\mrk{t} = t$. 
The least extension of the signature $\FS$ containing all such dependency pair symbols 
is denoted by $\FSs$. 
For a set $T \subseteq \TERMS$, we denote by $\mrk{T}$ the set of marked terms $\mrk{T} = \{ \mrk{t} \mid t \in T \}$. 
Let $\COM = \{\cs_0,\cs_1, \dots \}$ be a countable infinite set of fresh \emph{compound symbols}, 
where we suppose $\ar(\cs_n) = n$.
Compound symbols are used to group calls in \emph{dependency pairs for complexity} (\emph{dependency pairs} or \emph{DPs} for short).
We define $\com(t) = t$, and otherwise $\com(\seq{t}) = \cs_n(\seq{t})$ where $\cs_n \in \COM$. 
\begin{definition}[Dependency Pair, Dependency Pair Complexity Problem]\hfill
  \begin{enumerate}
  \item A \emph{dependency pair} (\emph{DP} for short) is a rewrite rule $\mrk{l} \to \com(\mrk{r_1},\dots,\mrk{r_n})$ 
    where $l,\seq{r} \in \TERMS$
    and $l$ is not a variable.
  \item Let $\SS$ and $\WW$ be two TRSs, 
    and let $\SSs$ and $\WWs$ be two sets of dependency pairs. 
    A complexity problem $\cp{\SSs \cup \SS}{\WWs \cup \WW}{\QQ}{\mrk{\TT}}$ with $\mrk{\TT} \subseteq \BTERMSs$ is called
    a \emph{dependency pair complexity problem} (or simply \emph{DP problem}).
  \end{enumerate}
\end{definition}
We keep the convention that $\RS,\SS,\WW,\dots$ are TRSs over $\TERMS$, whereas
$\RSs,\SSs,\WWs,\dots$ always denote sets of dependency pairs.

\begin{example}[Example~\ref{ex:pmult} continued]\label{ex:pmults}
  Denote by $\SSsmult$ the dependency pairs 
  \begin{align*}
    \rlbl{1}\colon\ \ms(x) \mrk{+} y & \to x \mrk{+} y & 
    \rlbl{2}\colon\ 0 \mrk{+} y & \to \cs_0\\
    \rlbl{3}\colon\ \ms(x) \mrk{\times} y & \to \cs_2(y \mrk{+} (x \times y), x \mrk{\times} y) & 
    \rlbl{4}\colon\ 0 \mrk{\times} y & \to \cs_0 \tkom
  \end{align*}
  and let $\Tbs$ be the (marked) basic terms with defined 
  symbols $\mrk{+},\mrk{\times}$ and constructors $\ms,0$. 
  Then $\PPmults \defsym \cp{\SSsmult}{\RSmult}{\RSmult}{\Tbs}$, 
  where $\RSmult$ are the rules for addition and multiplication depicted in 
  Example~\ref{ex:pmult}, is a DP problem. 
\end{example}
We anticipate that the DP problem $\PPmults$ reflects the complexity 
of our multiplication problem $\PPmult$, compare Theorem~\ref{rc:proc:tuples} below. 

For the remaining of this section, we fix a DP problem 
$\PPs = \cp{\SSs \cup \SS}{\WWs \cup \WW}{\QQ}{\TTs}$. 
Call an $n$-holed context $C$ a \emph{compound context} if it contains only compound symbols. 
Consider the $\PPmults$ derivation 
\begin{align*}
 % \underline{\ms(\ms(0)) \mrk{\times} \ms(0)}  
 % & \qrrew[\RSmult]{\SSsmult}{\RSmult} \cs_2(\underline{\ms(0) \mrk{+} \ms(0)}, \ms(0) \mrk{\times} \ms(0)) \\
 % & \qrrew[\RSmult]{\SSsmult}{\RSmult} \cs_2(0 \mrk{+} \ms(0), \underline{\ms(0) \mrk{\times} \ms(0)}) \\
 % & \qrrew[\RSmult]{\SSsmult}{\RSmult} \cs_2(0 \mrk{+} \ms(0), \cs_2(\ms(0) \mrk{+} (0 \times \ms(0)), 0 \mrk{\times} \ms(0))) \\
 % & \qrrew[\RSmult]{\SSsmult}{\RSmult} \cdots
 \underline{\ms(\ms(0)) \mrk{\times} \ms(0)}  
 & \rew[\PPmults] \cs_2( \ms(0) \mrk{+} (\text{\udensdash{$\ms(0) \times \ms(0)$}}), \ms(0) \mrk{\times} \ms(0)) \\
 & \rss[\PPmults] \cs_2(\underline{\ms(0) \mrk{+} \ms(0)}, \underline{\ms(0) \mrk{\times} \ms(0)}) \\
 & \rsl[\PPmults]{2} \cs_2(0 \mrk{+} \ms(0), \cs_2(\ms(0) \mrk{+} 0, 0 \mrk{\times} \ms(0))) 
 \tpkt
\end{align*}
Observe that any term in the above sequence can be written as $C[\seq{t}]$ where $C$ is a maximal compound 
context, and $\seq{t}$ are marked terms without compound symbols. 
For instance, the last term in this sequence 
is given as $C_1[0 \mrk{+} \ms(0), \ms(0) \mrk{+} 0, 0 \mrk{\times} \ms(0)]$ 
for $C_1 \defsym \cs_2(\hole, \cs_2(\hole, \hole))$. 
This holds even in general, but with the exception that $\seq{t}$ are not necessarily marked. 
Note that such an unmarked term $t_i$ ($i \in \{1,\dots,n\}$) can only result
from the application of a collapsing rules $\mrk{l} \to x$ for $x$ a variable, which 
is permitted by our formulation of dependency pair. 
We capture this observation with the set $\TERMSsfc$, defined
as the least extension of $\TERMS$ and $\TERMSs$ that is closed 
under compound contexts. Then the following observation holds.
\begin{lemma}\label{l:rc:compoundCtx}
  For every TRS $\RS$ and DPs $\RSs$, we have
  ${\rew[\RSs \cup \RS]}(\TERMSsfc) \subseteq \TERMSsfc$. 
  In particular,  $\fclosure{\mrk{\TT}}{\rew[\PPs]} \subseteq \TERMSsfc$ follows.
\end{lemma}
\begin{proof}
  Let $s = C[\seq{s}] \in \TERMSsfc$ where $C$ is a maximal compound context.
  Suppose $s \rew[\RSs \cup \RS] t$. Since $C$ contains only compound symbols, 
  it follows that 
  $t = C[s_1,\dots,t_i,\dots, s_n]$ where $s_i \rew[\RSs \cup \RS] t_i$ for some $i \in \{1,\dots,n\}$, 
  where again $t_i \in \TERMSsfc$.
  Conclusively $t \in \TERMSsfc$ and the first half of the lemma follows by inductive reasoning.
  From this the second half of the lemma follows, using that
  $\TTs \subseteq \TERMSsfc$ and taking $\RSs \defsym \SSs \cup \WWs$ and 
  $\RS \defsym \SS \cup \WW$. 
\end{proof}

Consider a term $t = C[\seq{t}] \in \TERMSsfc$ for a maximal compound context $C$. 
Any reduction of $t$ consists of
\emph{independent sub-derivations} of $t_i$ $(i = 1, \dots,n$), which are possibly interleaved.
To avoid reasoning up to permutations of rewrite steps, 
we introduce a notion of \emph{derivation tree} that disregards
the order of parallel steps under compound contexts. 
% For a dependency pair $\mrk{l} \to \cs_n(\mrk{r_1},\dots,\mrk{r_n}) \in \RSs$, let
% $s \leadsto_{l \to r} t$ hold if $s = \mrk{l}\sigma$ and 
% $t = \mrk{r_i}\sigma$ for some substitution $\sigma$.
\begin{definition}
  Let $t \in \TERMSs \cup \TERMS$. 
  The set of \emph{$\PPs$ derivation trees of $t$}, in notation $\DTree{\PPs}(t)$, 
  is defined as the least set of labeled trees such that:
  \begin{enumerate}
  \item $T \in \DTree{\PPs}(t)$ where $T$ consists of a unique node labeled by $t$. 
  \item Suppose $t \qrew[\QQ][\{l \to r\}] \com(\seq{t})$ for $l \to r \in \PPs$
    and let $T_i \in \DTree{\PPs}(t_i)$ for $i = 1,\dots,n$. 
    Then $T \in \DTree{\PPs}(t)$, where 
    $T$ is a tree with children $T_i$ ($i = 1,\dots,n$), 
    the root of $T$ is labeled by $t$, and the 
    edge from the root of $T$ to its children is labeled by $l \to r$.
  \end{enumerate}
  % The set of all $\PPs$ derivation trees is given by 
  % $$
  % \DTree{\PPs} \defsym \bigcup_{t \in \TT} \DTree{\PPs}(t) \tpkt
  % $$
\end{definition}

% \begin{example}[Example~\ref{ex:pmults} continued]\label{ex:pmults:tree}
%   In Figure~\ref{fig:pmults:tree} we depict a $\PPmults$ derivation tree of 
%   $\ms(\ms(0)) \mrk{\times} \ms(0)$.
%   % \newcommand{\el}[1]{edge from parent node {#1}}
% \end{example}
\begin{figure}
  \centering
  \vspace{-12pt}
  \newcommand{\trl}[1]{\scriptsize{#1}}
  \newcommand{\tnde}[1]{\footnotesize{#1}}
  \tikzstyle{nde}=[draw,fill=gray!10!white,solid,level distance=6mm, solid, inner sep=2pt, rounded corners=1mm]
  \tikzstyle{rl}=[level distance=6mm, inner sep=0pt,solid]
    \begin{tikzpicture}
      [ level 1/.style={sibling distance=38mm}
      , level 2/.style={sibling distance=40mm}
      , level 4/.style={sibling distance=28mm}]
      \node[nde] {\tnde{$\ms(\ms(0)) \mrk{\times} \ms(0)$}}
      child[rl] { node[rl] {\trl{$\ms(x) \mrk{\times} y  \to \cs_2(y \mrk{+} (x \times y), x \mrk{\times} y)$}}
        child[nde] { node[nde] {\tnde{$\ms(0) \mrk{+} (\ms(0) \times \ms(0))$}} %{\tnde{$\ms(0) \mrk{+} (\text{\udensdash{$\ms(0) \times \ms(0)$}})$}}
          child[rl] { node[rl] {\trl{$\ms(x) \times y \to y + (x \times y)$}}
            child[nde] { node[nde,yshift=-3mm] {\tnde{$\ms(0) \mrk{+} \ms(0)$}} edge from parent [dotted]  %edge from parent node [midway, sloped,fill=white] {$\cdots$}
              child[rl] { node[rl] {\trl{$\ms(x) \mrk{+} y \to x \mrk{+} y$}}
                child[nde] { node[nde] {\tnde{$0 \mrk{+} \ms(0)$}}}}}}}
        child[nde] { node[nde] {\tnde{$\ms(0) \mrk{\times} \ms(0)$}}
          child[rl] { node[rl] {\trl{$\ms(x) \mrk{\times} y  \to \cs_2(y \mrk{+} (x \times y), x \mrk{\times} y)$}}
            child[nde] { node[nde] {\tnde{$\ms(0) \mrk{+} 0$}}}
            child[nde] { node[nde] {\tnde{$0 \mrk{\times} \ms(0)$}}}}}}
      ; 
    \end{tikzpicture}
  \caption{$\PPmults$ Derivation Tree of $\ms(\ms(0)) \mrk{\times} \ms(0)$.}
  \label{fig:pmults:tree}
  \vspace{-12pt}
\end{figure}
Consider a $\PPs$ derivation tree $T$.
Note that an edge $e = \tuple{u,\sseq{v}}$ in $T$ precisely corresponds to a $\PPs$ step, 
in the sense that if $u$ is labeled by a term $t$ and $v_i$ ($i = 1,\dots,n$) by $t_i$, 
then $u \qrew[\QQ][l \to r] \com(\seq{t})$ holds, with $l \to r \in \PPs$ the label of $e$. 
In this case, we also say that the rule $l \to r$ was \emph{applied} at node $u$ in $T$.  
It is not difficult to see that from $T$ one can always extract a $\PPs$ derivation $D$, 
by successively applying the rewrite rules in $T$, starting from the root. 
In this case we also say that $D$ \emph{corresponds} to $T$, and vice versa. 
For instance, the derivation tree given in Figure~\ref{fig:pmults:tree}
and the derivation given below Example~\ref{ex:pmults} are corresponding.
Inversely, we can also associate to every $\PPs$ derivation $D$ starting from $t \in \TERMS \cup \TERMSs$ 
a $\PPs$ derivation tree $T$ corresponding to $D$, so that there is a one-to-one correspondence 
between applied rules in $T$ and rules applied in $D$.
This leads to following characterisation of the complexity function of $\PPs$.
Here $\size[\RSs \cup \RS]{T}$ refers to the number of applications of a rule $l \to r \in \RSs \cup \RS$ in $T$, 
more precisely, $\size[\RSs \cup \RS]{T}$ is the number of edges in 
$T$ labeled by a rule  $l \to r \in \RSs \cup \RS$. 

\begin{lemma}\label{l:rc:dt}
  For every $t \in \TERMS \cup \TERMSs$, we have 
  $$
  \dheight(t,\qrrew[\QQ]{\SSs \cup \SS}{\WWs \cup \WW}) \keq \max\{\size[\SSs \cup \SS]{T} \mid \text{$T$ is a $\PPs$-derivation tree of $t$}\} \tpkt
  $$
  In particular 
  $
  \cc[\PPs](n) \keq \max\{\size[\SSs \cup \SS]{T} \mid 
  \text{$T$ is a $\PPs$-derivation tree of $t \in \TT$ with $\size{t} \leqslant n$}\}
  $ holds. 
\end{lemma}
\begin{proof}
    For a term $t$, abbreviate
  $
  \max\{\size[\SSs \cup \SS]{T} \mid \text{$T$ is a $\PPs$-derivation tree of $t$}\}
  $ as $S$.
  Suppose $S$ is well-defined, 
  and let $T$ be a $\PPs$ derivation tree with $\size[\SSs \cup \SS]{T} = S$. 
  Without loss of generality, $T$ is finite. Otherwise we obtain a finite tree $T'$, 
  by removing from $T$ maximal sub-trees that contain only $\WWs \cup \WW$ nodes. 
  Then $\size[\SSs \cup \SS]{T} = \size[\SSs \cup \SS]{T'}$. Since by construction leafs 
  are targets of $\SSs \cup \SS$ edges, and the number of such edges is by assumption finite, 
  we see that the finitely branching tree $T'$ is of finite size.
  Since $T$ is a finite $\PPs$ derivation tree, a straight forward induction on $S$
  gives a $\qrrew[\QQ]{\SSs \cup \SS}{\WWs \cup \WW}$ derivation starting in $t$ of length $\size[\SSs \cup \SS]{T}$. 
  Hence $\dheight(t,\qrrew[\QQ]{\SSs \cup \SS}{\WWs \cup \WW}) \geqslant S$ whenever $\dheight(t,\qrrew[\QQ]{\SSs \cup \SS}{\WWs \cup \WW})$ is defined. 
  
  For the inverse direction, suppose $\ell = \dheight(t,\qrrew[\QQ]{\SSs \cup \SS}{\WWs \cup \WW}) \in \N$, and 
  consider a maximal derivation 
  $
  D : t = t_0 \qrrew[\QQ]{\SSs \cup \SS}{\WWs \cup \WW} t_1 \qrrew[\QQ]{\SSs \cup \SS}{\WWs \cup \WW} \dots \qrrew[\QQ]{\SSs \cup \SS}{\WWs \cup \WW} t_\ell
  $. 
  By induction on the length of the underlying $\rew[\PP]$ derivation it is not difficult to 
  construct a $\PPs$ derivation tree that witnesses 
  $S \geqslant \dheight(t,\qrrew[\QQ]{\SSs \cup \SS}{\WWs \cup \WW})$ (for $S$ defined). 
\end{proof}
\subsection{Weak Dependency Pairs and Dependency Tuples}

% The next definition introduces weak dependency pairs that can be used 
% both for runtime and innermost runtime complexity analysis. 
\begin{definition}[Weak Dependency Pairs~\cite{HM:IC:12}]
  Let $\RS$ denote a TRS such that the defined symbols of $\RS$ are included in $\DS$.  
  Consider a rule $l \to C[\seq{r}]$ in $\RS$, where $C$ is a maximal context containing 
  only constructors. 
  The dependency pair $\mrk{l} \to \com(\mrk{r_1},\dots,\mrk{r_n})$ is called a \emph{weak dependency pair}
  of $\RS$, in notation $\WDP(l \to r)$. 
  We denote by 
  $
   \WDP(\RS) \defsym \{ \WDP(l \to r) \mid l \to r \in \RS \}
  $
  the set of all weak dependency pairs of $\RS$. 
\end{definition}

In~\cite{HM:IJCAR:08} it has been shown that for any term $t \in \TERMS$, 
$\dheight(t,\rew[\RS]) = \dheight(\mrk{t},\rew[\WDP(\RS) \cup \RS])$. 
We extend this result to our setting, where the following lemma serves 
as a preparatory step. 
\begin{lemma}\label{l:rc:wdp}
  Let $\RS$ and $\QQ$ be two TRSs, such that the defined symbols of $\RS$ are included in $\DS$. 
  Then every derivation 
  $$
   t = t_0 \qrew[\QQ][\RS] t_1 \qrew[\QQ][\RS] t_2 \qrew[\QQ][\RS] \cdots \tkom
  $$
  for basic term $t$ is simulated step-wise by a derivation
  $$
   \mrk{t} = s_0 \qrew[\QQ][\WDP(\RS) \cup \RS] s_1 \qrew[\QQ][\WDP(\RS) \cup \RS]  s_2 \qrew[\QQ][\WDP(\RS) \cup \RS] \cdots \tkom
  $$
  and vice versa.
\end{lemma}
\begin{proof}
    For $\RS$ and $\QQ$ two TRSs, such that the defined symbols of $\RS$ are included in $\DS$, 
  we have to show that every derivation
  $$
   t = t_0 \qrew[\QQ][\RS] t_1 \qrew[\QQ][\RS] t_2 \qrew[\QQ][\RS] \cdots \tkom
  $$
  is simulated step-wise by a derivation
  $$
   \mrk{t} = s_0 \qrew[\QQ][\WDP(\RS) \cup \RS] s_1 \qrew[\QQ][\WDP(\RS) \cup \RS]  s_2 \qrew[\QQ][\WDP(\RS) \cup \RS] \cdots \tkom
  $$
  and vice versa.

  For a term $s$, let $P(s) \subseteq \Pos[\DS \cup \VS](s)$ be the set of minimal 
  positions such that the root of $s$ is in $\DS \cup \VS$. Hence in particular all positions in $P(s)$ are parallel. 
  Call a term $u = C[s_1, \dots, s_n]$ \emph{good for} $s$
  if $C$ is a context containing only constructors and compound symbols, 
  and there exists an injective mapping $\ofdom{m}{P(s) \to \Pos[\{\hole\}](C)}$ such that for all $p \in P(s)$, 
  $\subtermAt{u}{m(p)} = \subtermAt{s}{p}$ or $\subtermAt{u}{m(p)} = \mrk{(\subtermAt{s}{p})}$
  holds. Note that the mapping $m$ ensures that to every $\RS$ redex $\subtermAt{s}{p}$ we can associate 
  a possibly marked $\WDP(\RS) \cup \RS$ redex $\subtermAt{u}{m(p)}$. 
  % In other words, to each subterm $\subtermAt{s}{p}$ ($p \in P(s)$) 
  % we can associate a distinct occurrence $s_i$ ($i \in \{1,\dots,n\}$) in $u$ such that 
  % $s_i$ equals $\subtermAt{s}{p}$ or its marked version.

  Consider $s \qrew[\QQ][l \to r,p] t$ for $l \to r \in \RS$,  
  and suppose $\WDP(l \to r) = \mrk{l} \to \com(\mrk{r_1}, \dots, \mrk{r_m})$. 
  We show that for every term $u$ good for $s \in \TERMS$, there exists 
  a term $v$ with 
  $$u \qrew[\QQ][\{\WDP(l \to r), l \to r\}] v\tkom$$
  that is good for $t$. 
  This establishes the simulation from left to right. 
  Suppose $u = C[\seq{s}]$ is good for $s$ as witnessed by the mapping $\ofdom{m}{P(s) \to \Pos[\{\hole\}](C)}$
  and $C$ of the required form.
  Let $p'$ be a prefix of the rewrite position $p$ with $p' \in P(s)$. 
  This position exists, as the root of $\subtermAt{s}{p}$ is defined. 
  Let $s_i = \subtermAt{u}{m(p')}$ be the possible marked occurrence of $\subtermAt{s}{p'}$ in $u$. 
  We distinguish three cases. 

  Consider first the case $p' < p$. Then $\subtermAt{s}{p'} \qrew[\QQ][l \to r, {>}\varepsilon] \subtermAt{t}{p'}$ 
  by assumption. The latter implies
  $$
  u = C[s_1,\dots,s_i, \dots,s_n] \qrew[\QQ][l \to r] C[s_1,\dots,t_i, \dots,s_n] \symdef v \tkom
  $$
  for $s_i$ and $t_i$ the possibly marked versions of $\subtermAt{s}{p'}$ and $\subtermAt{t}{p'}$ respectively.
  Since by assumption $p' \in P(s)$ the root of $\subtermAt{s}{p'}$ and thus $\subtermAt{t}{p'}$
  is defined, it is not difficult to see that $P(s) = P(t)$ and $\ofdom{m}{P(s) \to \Pos[\{\hole\}](C)}$ witnessing 
  that $v$ is good for $t$. 

  Next consider that $p' = p$ and $s_i = \subtermAt{s}{p}$ is not marked, 
  by assumption thus $s_i = l\sigma$ for $\sigma$ a substitution such arguments of $l\sigma$ are $\QQ$ normal forms. 
  Conclusively
  $$
  u = C[s_1,\dots,l\sigma, \dots,s_n] \qrew[\QQ][l \to r] C[s_1,\dots,r\sigma, \dots,s_n] \symdef v \tpkt
  $$
  We claim $v$ is good for $t$. 
  Let $P(r\sigma) = \sseq[k]{q}$ and denote by $C_r$ the context 
  of $r\sigma$ with holes at positions $P(r\sigma)$.
  Set $C' \defsym C[\hole, \dots,C_r[\hole,\dots,\hole],\dots,\hole]$ 
  such that $v = C'[s_1,\dots,\subtermAt{r\sigma}{q_1},\dots,\subtermAt{r\sigma}{q_k},\dots,s_n]$,
  where in particular $C'$ contains only constructors or compound symbols. 
  Exploiting the mapping $\ofdom{m}{P(s) \to \Pos[\{\hole\}](C)}$ witnessing that $u$ is good for $s$, 
  it is not difficult to extend this to an injective function $\ofdom{m'}{P(t) \to \Pos[\{\hole\}](C')}$ 
  witnessing that $v$ is good for $t$:
  If $q \in P(t)$ is parallel to the rewrite position $p$, 
  we have $q \in P(s)$ and set $m'(q) \defsym m(q)$. 
  Note that $\subtermAt{v}{m'(q)} = \subtermAt{u}{m(q)}$ is some possibly marked 
  occurrence $s_j$ ($j = 1,\dots,n$) of $\subtermAt{s}{q} = \subtermAt{t}{q}$.
  For $q \in P(t)$ a position with $q = p \posc q'$ it follows that $q' \in P(r\sigma)$, 
  and we set $m'(q) = m(p) \posc q'$ where by construction $\subtermAt{v}{m'(q)} = \subtermAt{r\sigma}{q'} = \subtermAt{t}{q}$.
  This completes the definition of $m'$, as $q \in P(t)$ and $q < p$ 
  contradicts minimality of $p \in P(t)$. 

  The final case $p' = p$ but $\subtermAt{u}{m(p)}$ marked, i.e., $s_i = \mrk{(\subtermAt{s}{p})}$, is similar to above
  with the difference that we use the reduction
  $$
  u = C[s_1,\dots,s_i, \dots,s_n] \qrew[\QQ][\WDP(l \to r)] C[s_1,\dots,\com(\mrk{r_1}\sigma, \dots, \mrk{r_m}\sigma), \dots,s_n] \symdef v \tkom
  $$
  and for $C_r$ we use the maximal context of $\com(\mrk{r_1}\sigma, \dots, \mrk{r_m}\sigma)$ 
  containing only compound symbols or constructors. Here we exploit that $r_1\sigma,\dots,r_m\sigma$ contains 
  all occurrences of subterms of $\subtermAt{s}{p}$ that are variables or have a defined root symbol. 
  This completes the proof of the direction from left to right. 

  For the direction from right to left, 
  consider a term $u = C[\seq{s}]$ where $C$ is a compound context, and $s_i$ $(i = 1,\dots,n)$ possibly 
  marked terms without compound contexts. 
  Call a term $s \in \TERMS$ \emph{good for} $u$ if $s$ is obtained from $u$ by unmarking symbols, 
  and replacing $C$ with a context consisting only of constructors. 
  By case analysis on $u \qrew[\QQ][\WDP(\RSs) \cup \RS] v$, it can be verified that for any such $u$ 
  if $s$ is good for $u$, then there exists a term $t$ with $s \qrew[\QQ] t$ that is good for $v$. 
  Since the starting term $\mrk{t}$ is trivially of the considered shape, 
  the simulation follows. 
\end{proof}

\begin{theorem}[Weak Dependency Pair Processor]\label{rc:proc:wdp}
  Let $\PP = \cp{\SS}{\WW}{\QQ}{\TT}$ such 
  that all defined symbols in $\SS \cup \WW$ occur in $\DS$. 
  The following processor is sound and complete.
  $$
  \proc[Weak~Dependency~Pairs]{\judge{\cp{\SS}{\WW}{\QQ}{\TT}}{f}}{\judge{\cp{\WDP(\SS) \cup \SS}{\WDP(\WW) \cup \WW}{\QQ}{\mrk{\TT}}}{f}}
  $$
\end{theorem}
\begin{proof}
  Set $\PP \defsym \cp{\SS}{\WW}{\QQ}{\TT}$ and 
  $\PPs \defsym \cp{\WDP(\SS) \cup \SS}{\WDP(\WW) \cup \WW}{\QQ}{\mrk{\TT}}$.
  Suppose first $\cc[\PP] \in \bigO(f(n))$. 
  Lemma~\ref{l:rc:wdp} shows that every $\rew[\PP]$ reduction of $t \in \TT$
  is simulated by a corresponding $\rew[\PPs]$ reduction
  starting from $\mrk{t} \in \mrk{\TT}$. 
  Observe that every $\qrew[\QQ][\SS]$ step is simulated by a $\qrew[\QQ][\WDP(\SS) \cup \SS]$ step. 
  We thus obtain $\cc[\PPs] \in \bigO(f(n))$.
  This proves soundness, completeness is obtained dual. 
\end{proof}

% \begin{example}[Example~\ref{ex:pmult} continued]
%   The weak dependency pairs $\WDP(\RSmult)$ are given by 
%   \begin{align*}
%   \ms(x) \mrk{+} y & \to x \mrk{+} y 
%   & 0 \mrk{+} y & \to y &
%   \ms(x) \mrk{\times} y & \to y \mrk{+} (x \times y)
%   & 0 \mrk{\times} y & \to 0 \tpkt
%   \end{align*}
%   The weak dependency pair processor generates from $\PPmult$ the problem
%   $\cp{\WDP(\RSmult)}{\RSmult}{\RSmult}{\Tbs}$ for basic terms $\Tb$ as given in Example~\ref{ex:pmult}.
% \end{example}

We point out that unlike for termination analysis, 
to solve the generated sub-problems one
has to analyse applications of $\SS$ rules besides dependency pairs.
In contrast, DP problems of the form $\cp{\SSs}{\WWs \cup \WW}{\QQ}{\TTs}$
are often much easier to analyse.
In this situation rules that need to be accounted 
for, viz the strict rules, can only be applied in compound contexts.
Some processors tailored for DP problems 
can even only estimate the number of applications of dependency pair step, 
cf.\ for instance Theorem~\ref{rc:proc:rl}.
Notably, the strict order $\succ$ employed in a complexity pair $(\succsim,\succ)$
needs to be monotone only on compound contexts.
This is an immediate result of following observation. 
\begin{lemma}\label{l:rc:mucom}
  Let $\mu$ denote a usable replacement map for dependency pairs $\RSs$ in $\PPs$. 
  Then $\mucom$ is a usable replacement map for $\RSs$ in $\PPs$, 
  where $\mucom$ denotes the restriction of $\mu$ to compound symbols in the following sense:
  $\mucom(\cs_n) \defsym \mu(\cs_n)$ for 
  all $\cs_n \in \COM$, and otherwise $\mucom(f) \defsym \varnothing$ for $f \in \FSs$. 
\end{lemma}
\begin{proof}
  For a proof by contradiction, suppose $\mucom$ is not a usable replacement map for $\RSs$ in $\PP$. 
  Thus there exists $s \in \fclosure{\TT}{\rew[\PP]}$ 
  and position $p \in \Pos(s)$ such that $s \qrew[\QQ][\RSs,p] t$ for some term $t$, but $p \not \in \Pos[\mucom](s)$. 
  Since $s \in \TERMSsfc$ by Lemma~\ref{l:rc:compoundCtx}, symbols above position $p$ 
  in $s$ are compound symbols, and so $p \not \in \Pos[\mu](s)$ by definition of $\mucom$. 
  This contradicts however that $\mu$ is a usable replacment map for $\RSs$ in $\PP$. 
\end{proof}

We remark that using Lemma~\ref{l:rc:mucom} together with Theorem~\ref{rc:proc:rp}, 
our notion of $\PP$-monotone complexity pair generalises \emph{safe reduction pairs} from~\cite{HM:IJCAR:08},
that constitute of a rewrite preorder $\succsim$ and a total order $\succ$
stable under substitutions with ${\succsim} \cdot {\succ} \cdot {\succsim} \subseteq {\succ}$.
% In~\cite{HM:IJCAR:08,NEG:CADE:11}, safe reduction pairs are used to analyse dependency pairs modulo rewrite rules.%
% \footnote{Safe reduction pairs are called \emph{$\com$-monotone} in~\cite{NEG:CADE:11}.}  
It also generalises, theoretically,
the notion of \emph{$\mu$-monotone complexity pair} from~\cite{HM:IC:12}, 
that is parameterised by a single replacement map $\mu$ for all rules in $\PP$.%
\footnote{%
From a practical perspective, up to our knowledge only~$\TCT$ employs $\PP$-monotone
complexity pairs. \TCT\ however implements currently only the approximations presented in~\cite{HM:IC:12}.
} 

In~\cite{HM:IJCAR:08}, the \emph{weight gap principle} is introduced, with 
the objective to move the strict rules $\SS$ into the weak component, 
in order to obtain a DP problem of the form $\cp{\SSs}{\WWs \cup \WW}{\QQ}{\TTs}$, 
after the weak dependency pair transformation. 
\emph{Dependency tuples} introduced in~\cite{NEG:CADE:11} avoid the problem 
altogether. A complexity problem is directly translated into this form, 
at the expense of completeness and a more complicated set of dependency pairs. 

\begin{definition}[Dependency Tuples~\cite{NEG:CADE:11}]
  Let $\RS$ denote a TRS such that the defined symbols of $\RS$ are included in $\DS$.
  For a rewrite rule $l \to r \in \RS$, let $\seq{r}$ denote all subterms of the right-hand side 
  whose root symbol is in $\DS$. 
  The dependency pair $\mrk{l} \to \com(\mrk{r_1},\dots,\mrk{r_n})$ is called a \emph{dependency tuple}
  of $\RS$, in notation $\DT(l \to r)$. 
  We denote by 
  $
   \DT(\RS) \defsym \{ \DT(l \to r) \mid l \to r \in \RS \} %\tkom
  $,
  the set of all dependency tuples of $\RS$. 
\end{definition}

We generalise the central theorem from~\cite{NEG:CADE:11}, which shows that dependency tuples are sound 
for innermost runtime complexity analysis.
\begin{lemma}\label{l:rc:tuples}
  Let $\RS$ and $\QQ$ be two TRSs, such that the defined symbols of $\RS$ are included in $\DS$, 
  and such that $\NF(\QQ) \subseteq \NF(\RS)$.
  Then every derivation 
  $$
   t = t_0 \qrew[\QQ][\RS] t_1 \qrew[\QQ][\RS] t_2 \qrew[\QQ][\RS] \cdots \tkom
  $$
  for basic term $t$ is simulated step-wise by a derivation
  $$
   \mrk{t} = s_0 \qrrew[\QQ]{\DT(\RS)}{\RS} s_1 \qrrew[\QQ]{\DT(\RS)}{\RS}  s_2 \qrrew[\QQ]{\DT(\RS)}{\RS} \cdots \tpkt
  $$

\end{lemma}
\begin{proof}
    The proof follows the pattern of the proof of Lemma~\ref{l:rc:wdp}. 
  Define $P(s)$ as the restriction of $\Pos[\DS](s)$ that satisfies $\subtermAt{s}{p} \qrew[\QQ][\RS] u$ for each $p \in P(s)$ 
  and some term $u$. Observe that $P(s)$ contains in particular all redex positions in $s$. 
  Call a term $u = C[s_1, \dots, s_n]$ \emph{good for} $s$
  if $C$ is a context containing only constructors and compound symbols, 
  and there is some injective function $\ofdom{m}{P(s) \to \Pos[\{\hole\}](C)}$
  such that for every position $p \in \Pos(s)$, $\subtermAt{u}{m(p)} = \mrk{(\subtermAt{s}{p})}$.

  Consider a rewrite step $s = C[l\sigma] \qrew[\QQ][l \to r,p] C[r\sigma] = t$ for 
  position $p$, context $C$, substitution $\sigma$ and rewrite rule $l \to r \in \RS$. 
  Observe that $P(t) \subseteq (P(s) \setminus \{p\}) \cup \{ p\posc q \mid q \in \Pos[\DS](r)\}$. 
  For this, suppose $q \in P(t)$. 
  If $q \pospar p$ for the rewrite position $p$, then $\subtermAt{s}{p} = \subtermAt{t}{p}$ 
  and so $q \in P(s)$. 
  For $q < p$, observe that roots of $\subtermAt{s}{p}$ and $\subtermAt{t}{p}$ coincide, 
  in particular the assumption $q \in P(t)$ thus gives $q \in \Pos[\DS](s)$ and 
  the assumption $s \qrew[\QQ][\RS,p] t$ ensures that again $q \in P(s)$. 
  Finally consider $q > p$, that is $q = p \posc q'$ for some position $q' \in \Pos[\DS](r\sigma)$
  with $\subtermAt{r\sigma}{q'} \qrew[\QQ][\RS] u$ for some term $u$. 
  Note that by the assumption $\NF(\QQ) \subseteq \NF(\RS)$, for every variable $x$ in $r$, 
  $x\sigma \in \NF(\QQ) \subseteq \NF(\RS)$ holds. Conclusively $q' \in \Pos[\DS](r)$
  and the assertion follows again. 

  We now show that if $u = C[\seq{s}]$ is good for $s$, then $u \qrrew[\QQ]{\DT(\RS)}{\RS} v$ holds 
  for some term $v$ good for $t$. 
  Set $\mrk{l} \to \com(\mrk{r_1},\dots,\mrk{r_n}) \defsym \DT(l \to r)$. 
  Using that $p \in P(s)$,
  $$
    u = C[s_1,\dots,\mrk{l}\sigma, \dots,s_n] \qrew[\QQ][\DT(l \to r)] C[s_1,\dots,\com(\mrk{r_1}\sigma, \dots, \mrk{r_m}\sigma), \dots,s_n] \symdef v' \tkom
  $$
  holds. 
  Hence for $C' = C[\hole, \dots, \com(\hole,\dots,\hole), \dots, \hole]$, 
  $v' = C'[s_1,\dots,\mrk{r_1}\sigma, \dots, \mrk{r_m}\sigma, \dots,s_n]$. 
  We verify that $v' \qrss[\QQ][\RS] v$ for some $v$ good for $t$. 
  Recall that by the observation on $P(t)$, every 
  position $q \in P(t) \setminus P(s)$ can be decomposed 
   $q = p \posc q_i$ for some position $q_i \in \Pos[\DS](r)$, 
  with $\subtermAt{r}{q_i} = r_i$ for $i = 1,\dots,m$. 
  Let $q_i'$ denote the position of the occurrence $\mrk{r_i}$ in 
  the right-hand side $\com(\mrk{r_1}, \dots, \mrk{r_m})$, 
  and set $m(q) \defsym m(p) \posc q_i'$. 
  Note that the resulting function is an injective function from $P(t)$ to $\Pos[\{\hole\}](C')$.
  By construction we have $\subtermAt{v'}{m(q)} = \mrk{r_i}\sigma = \mrk{(\subtermAt{t}{q})}$ for all 
  positions $q = p \posc q_i \in P(t) \setminus P(s)$. 
  For $q$ not of this shape we have $q \in P(s) \setminus \{p\}$ by the observation on $P(t)$, 
  in particular either $\mrk{(\subtermAt{s}{q})} = \mrk{(\subtermAt{t}{q})}$ 
  or otherwise $\subtermAt{s}{q} \not = \subtermAt{t}{q}$ and the assumption $q \not = p$ gives $q < p$. 
  Conclusively $\mrk{\subtermAt{t}{q}} \qrss[\QQ][l \to r,{>}\posempty] \mrk{\subtermAt{t}{q}}$. 
  Hence rewriting in $v'$ all terms $\subtermAt{v'}{m(q)} = \mrk{(\subtermAt{s}{q})}$ 
  with $q \in P(s) \setminus \{p\}$ to $\mrk{\subtermAt{t}{q}}$ gives the desired term $v$ good for $t$.
\end{proof}

% The next theorem provides a generalisation of~\cite[Theorem~10]{NEG:CADE:11}.
\begin{theorem}[Dependency Tuple Processor]\label{rc:proc:tuples}
  Let $\PP = \cp{\SS}{\WW}{\QQ}{\TT}$ be an \emph{innermost} complexity problem
  such that all defined symbols in $\SS \cup \WW$ occur in $\DS$. 
  The following processor is sound.
  $$
  \proc[Dependency~Tuples]{\judge{\cp{\SS}{\WW}{\QQ}{\TT}}{f}}{\judge{\cp{\DT(\SS)}{\DT(\WW) \cup \SS \cup \WW}{\QQ}{\mrk{\TT}}}{f}}
  $$
\end{theorem}
\begin{proof}
  Reasoning identical to Theorem~\ref{rc:proc:wdp}, using Lemma~\ref{l:rc:tuples}.
\end{proof}
Note that the problem $\PPmults$ depicted in Example~\ref{ex:pmults} is generated by the 
dependency tuple processor.

\section{Dependency Pair Processors}\label{s:rc:dpprocs}

The dependency pair method opened the door for a wealth of powerful termination 
techniques. 
In the literature, the majority of these techniques have been 
suitably adapted to complexity analysis. 
% As explained before,\emph{ (safe) reduction pairs}, and our generalisation of 
% $\PP$-monotone complexity pairs, are admissible for runtime complexity analysis.
For instance, in~\cite{HM:IJCAR:08} it is shows that \emph{usable rules} are sound 
for runtime complexity analysis.
\emph{Cycle analysis}~\cite{HM:IC:05} on the other hand is not sound in general, 
but \emph{path analysis}~\cite{HM:IJCAR:08} constitutes an adaption of this technique
for complexity analysis. Both techniques can be easily adapted to our setting. 
For innermost rewriting in conjunction with dependency tuples, 
in~\cite{NEG:CADE:11} the processors based on pair transformations~\cite{T:07} are proven sound. 
Noteworthy, some techniques have recently been establishes directly for 
DP complexity problems. 
For instance, the aforementioned \emph{weight gap principle}~\cite{HM:IJCAR:08,HM:IC:12}
and the \emph{remove leafs} and \emph{knowledge propagation} processor from~\cite{NEG:CADE:11}.%
\footnote{The weight gap principle was later adapted by~\cite{ZK:RTA:10} to their setting.}
Except for the latter two processors, adapting the above mentioned techniques to our setting 
is an easy exercise. Due to the presence of weak rules in our notion of complexity problem, 
the remove leafs processor is even unsound. Still, the combination of 
the two processors presented in Theorem~\ref{rc:proc:rl} and Theorem~\ref{rc:proc:pe}
allow a simulation, where sound. 

The first processor we want to discuss stems from a careful
analysis of the \emph{dependency graph}.
Throughout the following, we fix again a DP problem $\PPs = \cp{\SSs \cup \SS}{\WWs \cup \WW}{\QQ}{\TTs}$.
\begin{definition}[Dependency Graph]
  The nodes of the \emph{dependency graph} (\emph{DG} for short) $\GS$ of $\PPs$ are the 
  dependency pairs from $\SSs \cup \WWs$, and there is an arrow labeled by $i \in \N$ from 
  $\mrk{s} \to \com(\mrk{t_1},\dots,\mrk{t_n})$ to $\mrk{u} \to \com(\mrk{v_1},\dots,\mrk{v_m})$ 
  if for some substitutions $\ofdom{\sigma,\tau}{\VS \to \TERMS}$, 
  $\mrk{t_i}\sigma \qrss[\QQ][\SS \cup \WW] \mrk{u}\tau$.
\end{definition}
\begin{figure}
  \vspace{-12pt}
  \begin{center}
    \begin{tikzpicture}
      \node[dgnde]                   (m1) {\strut \rlbl{3}};
      \node[dgnde,below of=m1]       (m2) {\strut \rlbl{4}};
      \node[dgnde,below right of=m1] (a1) {\strut \rlbl{1}};
      \node[dgnde,below of=a1]       (a2) {\strut \rlbl{2}};

      \path[->] (m1) edge [loop right]  node [midway, above] {2} (m1);
      \path[->] (a1) edge [loop right]  node [midway, above] {1} (a1);
      \path[->] (m1) edge node [midway, above right, inner sep =0mm] {1} (a1);
      \path[->] (m1) edge node [midway, right, inner sep = 0mm] {2} (m2);
      \path[->] (a1) edge node [midway, right, inner sep = 0mm] {1} (a2);
    \end{tikzpicture}
    \caption{DG of $\PPmults$.}
    \label{fig:pmults:dg}
    \vspace{-12pt}
  \end{center}
\end{figure}
Usually the weak dependency graph of $\PPs$ is not computable.
We say that $\GS$ is a \emph{weak dependency graph approximation} 
for $\PPs$ if $\GS$ contains the DG of $\PPs$.
Figure~\ref{fig:pmults:dg} depicts the dependency graph of $\PPmults$, 
where \rlbl{1} --- \rlbl{4} refer to the DPs given in Example~\ref{ex:pmults}.
The dependency graph $\GS$ tells us in which order dependency pairs can occur 
in a derivation tree of $\PPs$. To make this intuition precise, 
we adapt the notion of \emph{DP chain} known from termination analysis to derivation trees.
Recall that for derivation tree $T$, $\dtsucc[T]$ denotes the successor relation, 
and $\dtsucc[T][l \to r]$ its restriction to edges with applied rule $l \to r$. 
\begin{definition}[Dependency Pair Chain]
  Consider a $\PPs$ derivation tree $T$ and nodes $u_1,u_2,\dots$. 
  such that 
  $$
  u_1 \dtsucc[T][l_1 \to r_1] \cdot \dtsucc[T][\SS \cup \WW]^{*}
  u_2 \dtsucc[T][l_2 \to r_2] \cdot \dtsucc[T][\SS \cup \WW]^{*}
  \cdots \tkom
  $$
  holds for dependency pairs $C\colon l_1 \to r_1,l_2 \to r_2, \dots$.
  The sequence $C$ 
  is called a \emph{dependency pair chain (in $T$)}, or \emph{chain} for brevity.
\end{definition}
The next lemma is immediate from the definition. 
\begin{lemma}\label{l:rc:chain}
  Every chain in a $\PPs$ derivation tree is a path in the dependency graph of $\PPs$. 
  % Let $\GS$ be a DG approximation for $\PPs$. 
  % Then every chain in a $\PPs$ derivation tree $T$ is a path in $\GS$.
\end{lemma}
\begin{proof}
  Let $\PPs = \cp{\SSs \cup \SS}{\WWs \cup \WW}{\QQ}{\TTs}$ and consider two successive elements 
  $l_1 \to r_1 \defsym \mrk{s} \to \com(\mrk{t_1},\dots,\mrk{t_n})$ and 
  $l_2 \to r_2 \defsym \mrk{u} \to \cs_m(\mrk{v_1},\dots,\mrk{v_m})$ in a dependency pair chain 
  of a $\PPs$ derivation tree $T$. 
  Thus there exists nodes $u_1,u_2,v_1,v_2$ with
  $$
  u_1 \dtsucc[T][l_1 \to r_1] 
  u_2 \dtsucc[T][\SS \cup \WW]^{*}
  v_1 \dtsucc[T][l_2 \to r_2]
  v_2 \tkom
  $$
  and thus there exists substitutions $\sigma,\tau$ such that $u_2$ 
  is labeled by $\mrk{t_i}\sigma$ for some $i \in \{1,\dots,n\}$
  and $v_1$ by $\mrk{u}\tau$. As $u_2 \dtsucc[T][\SS \cup \WW]^{*} v_1$ we have 
  $\mrk{t_i}\sigma \qrss[\QQ][\SS \cup \WW] \mrk{u}\tau$ by definition, 
  and thus there is an edge from $l_1 \to r_1$ to $l_2 \to r_2$ in the WDG of $\PPs$, 
  and hence in $\GS$.
  The lemma follows from this.
\end{proof}

Denote by $\Pre[\GS](l \to r)$ the set of all (direct) \emph{predecessors} of node $l \to r$ in $\GS$, 
for a set of dependency pairs $\RSs$ we set $\Pre[\GS](\RSs) \defsym \bigcup_{l \to r \in \RSs} \{\Pre[\GS](l \to r)\}$. 
Noschinski~et~al.~\cite{NEG:CADE:11} observed that the application of a dependency 
pair $l \to r$ in a $\PPs$ derivation can be estimated in terms of the application of its predecessors 
in the dependency graph of $\PPs$. 
For this note that any application of $l \to r$ in a $\PPs$ derivation 
tree $T$ of $\mrk{t} \in \TTs$ is either at the root, or by Lemma~\ref{l:rc:chain} preceded by the application 
of a predecessor $l' \to \com(\seq{r})$ of $l \to r$ in the dependency graph of $\PPs$. 
Precisely, we have following correspondence, where $K$ is used to approximate $n$. 
\begin{lemma}\label{l:rc:pre}
  Let $\GS$ be an approximated dependency graph for $\PPs$. 
  For every $\PPs$-derivation tree $T$,   
  $\size[\RSs \cup \RS]{T} \leqslant \max\{1,\size[(\RSs \setminus \{l \to r\}) \cup {\Pre[\GS](l \to r)}  \cup \RS]{T} \cdot K \}$
  where $K$ denote the maximal arity of a compound symbol in $\PPs$.
\end{lemma}
\begin{proof}
  Consider the non-trivial case $l \to r \not\in \Pre[\GS](l \to r)$ and let $T$ denote a $\PPs$ derivation tree
  with an edge labeled by $l \to r \in \RSs$. 
  It suffices to verify $\size[\{l \to r\}]{T} \leqslant \max\{1,\size[{\Pre[\GS](l \to r)}]{T} \cdot K\}$.
  By Lemma~\ref{l:rc:chain} chains of $T$ translate to paths in $\GS$, and consequently
  if $l \to r$ occurs in $T$, then the assumption $l \to r \not\in \Pre[\GS](l \to r)$
  gives that either $l \to r$ occurs only in the beginning of chains, or 
  is headed by a dependency pair from $\Pre[\GS](l \to r)$. 
  In the former case $\size[\{l \to r\}]{T} = 1$. 
  In the latter case, let $\{\seq{u}\}$ collect all sources of $l \to r$ edges in $T$.
  To each node $u_i \in \{\seq{u}\}$ we can identify a unique node $\m{pre}(u_i)$ such that 
  $\m{pre}(u_i) \dtsucc[T][{\Pre[\GS](l \to r)}] \cdot \dtsucc[T][\SS \cup \WW]^* u_i$.
  Let $\{\seq[m]{v}\} = \{\m{pre}(u_1), \dots, \m{pre}(u_n)\}$.
  Since $\dtsucc[T][\SS \cup \WW]$ is non-branching, and $\m{pre}(u_i)$
  has at most $K$ successors, it follows 
  that 
  $\size[\{l \to r\}]{T} = n \leqslant K \cdot m \leqslant K \cdot \size[{\Pre[\GS](l \to r)}]{T}$.
\end{proof}
This observation gives rise to following processor.
\begin{theorem}[Predecessors Estimation]\label{rc:proc:pe}
  Let $\GS$ be an approximated dependency pair graph of $\PPs$.
  The following processor is sound:
  $$
  \proc[Predecessor~Estimation]
       {\judge{\cp{\SSs_1 \cup \SSs_2 \cup \SS}{\WWs \cup \WW}{\QQ}{\TTs}}{f}}
       {\judge{\cp{\Pre[\GS](\SSs_1) \cup \SSs_2 \cup \SS}{\SSs_1 \cup \WWs \cup \WW}{\QQ}{\TTs}}{f}}
       \tpkt
  $$
\end{theorem}
\begin{proof}
  The Lemma follows from Lemma~\ref{l:rc:pre} and Lemma~\ref{l:rc:dt}.
\end{proof}

We point out that the predecessor estimation processor is an adaption of 
\emph{knowledge propagation} introduced in~\cite{NEG:CADE:11}. 
The notion of problem from~\cite{NEG:CADE:11} uses for this processor specifically 
a dedicated component $\mathcal{K}$ of rules with known complexity, 
and $l \to r$ can be move to this component if $\mathcal{K}$ contains all predecessors 
of $l \to r$. Although we could in principle introduce such a component in our notion, 
we prefer our formulation of the predecessor estimation that does not rely on $\mathcal{K}$.
\begin{example}[Example~\ref{ex:pmults} continued]\label{ex:pmultss}
  Reconsider the dependency graph $\GS$ of $\PPmults$ given in Figure~\ref{fig:pmults:dg}. 
  The predecessor estimation processor allows us to estimate the number of applications
  of rules $\{\rlbl{2},\rlbl{4}\}$ in terms of their 
  predecessors $\Pre[\GS](\{\rlbl{2},\rlbl{4}\}) = \{\rlbl{1},\rlbl{3}\}$ as follows:
  $$
  \proc[Predecessor~Estimation]
       {\judge{\cp{\{\rlbl{1},\rlbl{2},\rlbl{3},\rlbl{4}\}}{\RSmult}{\RSmult}{\Tbs}}{f}}
       {\judge{\cp{\{\rlbl{1},\rlbl{3}\}}{\{\rlbl{2},\rlbl{4}\} \cup \RSmult}{\RSmult}{\Tbs}}{f}}
       \tpkt
  $$
\end{example}

The \emph{remove leafs} processor~\cite{NEG:CADE:11} states that all leafs from 
the dependency graph can be safely removed. This processor is unsound in the presence of 
weak dependency pairs $\WWs$. It is not difficult to see that the complexity of
$\cp{\{\m{\mrk{g}} \to \mc_0\}}{\{\m{\mrk{f}} \to \cs_2(\m{\mrk{f}},\m{\mrk{g}})\}}{\varnothing}{\{\m{\mrk{f}}\}}$ 
is undefined, whereas the complexity of the problem 
$\cp{\varnothing}{\{\m{\mrk{f}} \to \cs_2(\m{\mrk{f}},\m{\mrk{g}})\}}{\varnothing}{\{\m{\mrk{f}}\}}$,
obtained by removing $\m{\mrk{g}} \to \mc_0$ that occurs as leaf in the dependency graph, 
is constant. 
The problem $\cp{\{\m{\mrk{f}} \to \cs_1(\m{g}),~\m{g} \to \m{g}\}}{\varnothing}{\varnothing}{\{\m{\mrk{f}}\}}$ 
witnesses that also in the presence of a non-empty set $\SS$ of strict rules, 
this processor is unsafe. 
Provided that $\SS = \varnothing$, we can however remove leafs from the dependency graph 
that do not belong to $\SSs$. This observation can be generalised, as captured in the following processor.
Below we denote by $\cutAt{T}{\RS}$, for a $\PP$ derivation tree $T$ and rewrite system $\RS$, the \emph{trim} 
of $T$ to $\RS$ nodes, obtained by removing sub-trees in $T$ not rooted at $\RS$ nodes. 
More precise, $\cutAt{T}{\RS}$ denotes the sub-graph of 
$T$ accessible from the root of $T$, after removing all edges \emph{not} labeled by rules 
from $\RS$. Note that $\cutAt{T}{\RS}$ is again a $\PP$ derivation tree.

\begin{theorem}[Remove Weak Suffix Processor]\label{rc:proc:rl}
  Let $\GS$ be an approximated dependency graph of $\PPs = \cp{\SSs}{\WWs_1 \cup \WWs_2 \cup \WW}{\QQ}{\TTs}$, 
  where $\WWs_1$ is closed under $\GS$ successors. 
  $$
  \proc[Remove~Weak~Suffix]{\judge{\cp{\SSs}{\WWs_1 \cup \WWs_2 \cup \WW}{\QQ}{\TTs}}{f}}
       {\judge{\cp{\SSs}{\WWs_2 \cup \WW}{\QQ}{\TTs}}{f}}  \tpkt
  $$
\end{theorem}
\begin{proof}
  Let $\PPs = \cp{\SSs}{\WW \cup \WW}{\QQ}{\TT}$,
  and $\PP_\uparrow = \cp{\SSs}{\WWs_\uparrow \cup \WW}{\QQ}{\TT}$
  and consider $t \in \TT$. 
  For soundness, suppose $\judge{\PP_\uparrow}{f}$ is valid.
  Consider a $\PPs$ derivation tree $T$ of $t \in \TT$. 
  Then $T_\uparrow \defsym \cutAt{T}{\SSs \cup \WWs_\uparrow \cup \WW}$ is a $\PP_\uparrow$ derivation 
  tree. Note that $\size[\SSs](T_\uparrow) = \dheight(t,\rew[\PP_\uparrow])$ 
  by Lemma~\ref{l:rc:dt} is well defined. 
  We claim $\size[\PP_\uparrow]{T_\uparrow} = \size[\PPs]{T}$. 
  Otherwise $\size[\PPs]{T}$ is either not defined or 
  $\size[\PPs]{T} > \size[\PP_\uparrow]{T_\uparrow}$. 
  Any case implies that there is some path 
  $$
  u_1 \dtsucc[T][l \to r] \cdot \dtsucc[T]^* u_2 \dtsucc[T][l' \to r'] \tkom
  $$
  with $l' \to r' \in \SSs$ but $u_1$ is a leaf in $T_\uparrow$. 
  Together with Lemma~\ref{l:rc:chain} this however contradicts the assumption on $l \to r$.  
  Since $T$ was arbitrary, Lemma~\ref{l:rc:dt} proves that 
  $\judge{\PPs}{f}$ is valid, we conclude soundness. 

  For the inverse direction, observe that by definition any $\PP_\uparrow$ derivation 
  tree is also a $\PPs$ derivation tree. Completeness thus follows by Lemma~\ref{l:rc:dt}.
\end{proof}

\begin{example}[Example~\ref{ex:pmultss} continued]\label{ex:pmultsss}
  The above processor finally allows us to delete the leafs $\rlbl{2}$ and $\rlbl{4}$ 
  from the sub-problem generated in Example~\ref{ex:pmultss}:
  $$
  \proc[Remove~Weak~Suffix]
    {\judge{\cp{\{\rlbl{1},\rlbl{3}\}}{\{\rlbl{2},\rlbl{4}\} \cup \RSmult}{\RSmult}{\Tbs}}{f}}
    {\judge{\cp{\{\rlbl{1},\rlbl{3}\}}{\RSmult}{\RSmult}{\Tbs}}{f}}
       \tpkt
  $$
\end{example}

In the remining of this section, 
we focus on a novel technique that we call \emph{dependency graph decomposition}.
This technique is greatly motivated by 
the fact that none of the transformation processors from the cited literature 
is capable of translating a problem to computationally simpler 
sub-problems: any complexity proof is of the form
$\judge[\jdgmt{\PP_1}{f_1}, \dots, \jdgmt{\PP_n}{f_n}]{\PP}{f}$ 
for $f_i \in \bigO(f)$ ($i = 1,\dots,n$). 
From a modularity perspective, the processors introduced so far are 
to some extend disappointing. With exception of the last processor
that removes weak dependency pairs, 
none of the processors is capable of making the input problem smaller.
Worse, none of the processors allows the decomposition of a problem $\PPs$ into 
sub-problems with asymptotically strictly lower complexity.
This even holds for the decomposition processor given in Theorem~\ref{rc:proc:decompose}.
This implies that the maximal bound one can prove is 
essentially determined by the strength of the employed base techniques, 
viz complexity pairs.
In our experience however, a complexity prover
is seldom able to synthesise a suitable complexity pair that induces 
a complexity bound beyond a cubic polynomial. 
Notably small polynomial path orders~\cite{AEM:TCS:12} present, due to its syntactic nature,
an exception to this.

\begin{figure}
  \vspace{-12pt}
  \centering
  \begin{tikzpicture}
    \begin{scope}[scale=0.5,xshift=-3cm]
      
      \draw[draw,fill=black!20!white] (0,1) -- (-1cm,-0.5cm) -- (3.5cm,-1.5cm) -- (0,1);
      
      \node (S) at (0,1.1) {};
      \node (S) at (0.3,0.15) {$T_\uparrow$};
      
      \draw[thick,dotted] (-0.5cm,-1.4cm) -- (2.5cm,-2.06cm);
      
      \begin{scope}[xshift=-0.5cm,yshift=0cm]
        \node (T1) at (0,0) {};
        \draw[draw,fill=white] (0,0) -- (-0.5cm,-2cm) -- (0.5cm,-2cm) -- (0,0);
        % \node at (0,0.1) {$t_1$};
        \node at (0,-2.5) {$T_1$};
      \end{scope}
      
      \begin{scope}[xshift=1.cm,yshift=-0.45cm]
        \node (Ti) at (0,0) {};
        \draw[draw,fill=white] (0,0) -- (-0.5cm,-2cm) -- (0.5cm,-2cm) -- (0,0);
        % \node at (0,0.1) {$t_i$};
        \node at (0,-2.5) {$T_i$};
      \end{scope}
      
      \begin{scope}[xshift=2.5cm,yshift=-0.90cm]
        \node (Tn) at (0,0) {};
        \draw[draw,fill=white] (0,0) -- (-0.5cm,-2cm) -- (0.5cm,-2cm) -- (0,0);
        % \node at (0,0.1) {$t_n$};
        \node at (0,-2.5) {$T_n$};
      \end{scope}
      
      % \node at (1.25,-4.5cm) {Derivation Tree of $\ms^k(0) \mrk{\times} \ms^l(0)$};
    \end{scope}
  \end{tikzpicture}
  \vspace{-10pt}
  \caption{Upper and lower layer in $\PPs$-derivation tree $T$.}
  \label{fig:dt:sep}
  \vspace{-10pt}
\end{figure}
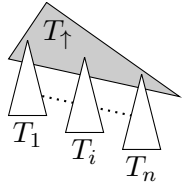

To keep the presentation simple, suppose momentarily that $\PPs$ is of the form $\cp{\SSs}{\WW}{\QQ}{\TTs}$. 
Consider a partitioning $\SSs_\downarrow \cup \SSs_\uparrow = \SSs$, 
and associate with this partitioning the two complexity problems
$\PPs_\downarrow \defsym \cp{\SSs_\downarrow}{\WW}{\QQ}{\TTs}$ and 
$\PPs_\uparrow \defsym \cp{\SSs_\uparrow}{\WW}{\QQ}{\TTs}$.
Suppose $\SSs_\downarrow \subseteq \SSs$ is \emph{forward closed}, that is, it is 
closed under successors with respect to the dependency graph of $\PPs$.
As depicted in Figure~\ref{fig:dt:sep}, the partitioning on dependency 
pairs also induces a partitioning on $\PPs$-derivation 
trees $T$ into two (possibly empty) layers:
the \emph{lower layer} constitutes of the \emph{maximal} subtrees $\seq{T}$ 
of $T$ that are $\PPs_\downarrow$-derivation trees;
the \emph{upper layer} is given by the tree $T_\uparrow$, obtained by removing from $T$ the subtrees $T_i$ ($i = 1,\dots,n$). 
Note that since $\SS_\downarrow$ is forward closed and the subtrees $T_i$ are maximal, 
Lemma~\ref{l:rc:chain} yields that all DPs applied in the upper layer $T_\uparrow$ occur in $\SSs_\uparrow$.

In order to bind $\size[\SSs]{T}$ as a function in the size of the initial term $t$, 
and conclusively the complexity of $\PPs$ in accordance to Lemma~\ref{l:rc:dt}, 
the decompose processor analyses the upper layer $T_\uparrow$ and the subtrees $T_i$ ($i = 1,\dots,n$)
from the lower layer separately. 
Since $\size[\SSs_\uparrow]{T_\uparrow}$ accounts for application of strict rules in the upper layer
\emph{and} the number $n$ of subtrees from the lower layer, 
it is tempting to think that two complexity proofs 
$\judge{\PPs_\uparrow}{f}$ and $\judge{\PPs_\downarrow}{g}$
verify $\size[\SSs]{T} \in \bigO(f(\size{t}) \cdot g(\size{t}))$. 
Observe however that the trees $T_i$ $(i = 1,\dots,n)$ are not necessarily derivation trees 
of terms from $\TTs$. The argument thus breaks since $g$ cannot bind 
the applications of strict rules in $T_i$ in general. For this, consider the following 
example.
\begin{example}
  Consider the TRS $\RSexp$ that expresses exponentiation. 
  \begin{align*}
    \rlbl{e}\colon\ \m{d}(0) & \to 0 & 
    \rlbl{f}\colon\ \m{d}(\ms(x)) & \to \m{s}(\m{s}(\m{d}(x))) & 
    \rlbl{g}\colon\ \m{e}(0) & \to \ms(0) & 
    \rlbl{h}\colon\ \m{e}(\ms(x)) & \to \m{d}(\m{e}(x)) \tpkt
  \end{align*}
  Using the dependency tuple processor and the above simplification processors,  
  it is not difficult to show that 
  $\judge{\cp{\RSexp}{\varnothing}{\RSexp}{\Tb}}{f}$ follows from 
  $\judge{\cp{\SSsexp}{\RSexp}{\RSexp}{\Tbs}}{f}$, where $\SSsexp$ and the DG are given by 
  \begin{align*}
    \rlbl{5}\colon\ \m{\mrk{d}}(\ms(x)) & \to \m{\mrk{d}}(x) & 
    \quad \rlbl{6}\colon\ \m{\mrk{e}}(\ms(x)) & \to \cs_2(\m{\mrk{e}}(x), \m{\mrk{d}}(\m{e}(x))) \tpkt
  \end{align*}
  The dependency graph consists of two cycles $\{\rlbl{6}\}$ and $\{\rlbl{5}\}$ that both admit linear complexity, that is, 
the complexity function of 
$\cp{\{\rlbl{5}\}}{\RSexp}{\RSexp}{\Tbs}$ and also
$\cp{\{\rlbl{6}\}}{\RSexp}{\RSexp}{\Tbs}$ is bounded by a linear polynomial. 
On the other hand, the complexity function of $\cp{\RSexp}{\varnothing}{\RSexp}{\Tb}$ 
is asymptotically bounded by an exponential from below.
\end{example}
The gap above is caused as the above decomposition into the two cycles does not account for the specific calls
from the upper components (cycle $\{\rlbl{6}\}$ in the above example), 
to the lower components (cycle $\{\rlbl{5}\}$). 
To rectify the situation, one could adopt the set of starting terms in $\PPs_\downarrow$.
In order to assure that starting terms of the obtained problem are basic, 
we instead add sufficiently many dependency pairs to the weak component of $\PPs_\downarrow$, 
that generate this set of starting terms accordingly.
Let $\m{sep}(\SSs_\uparrow)$ constitute of all DPs $l \to r_i$ for
$l \to \com(r_1,\dots,r_i, \dots,r_m) \in \SSs_\uparrow$. 
Together with weak rules $\WW$, the rules $\m{sep}(\SSs_\uparrow)$ are sufficient to 
simulate the paths from the root of $T_\uparrow$ to the subtrees $T_i$ ($i = 1,\dots,n$). 
A complexity proof $\judge{\cp{\SSs_\downarrow}{\m{sep}(\SSs_\uparrow) \cup \WW}{\QQ}{\TTs}}{g}$ 
thus verifies that application of strict rules in $T_i$ are bounded by  $\bigO(g(\size{t}))$ as desired. 

We demonstrate this decomposition on our running example. 
\begin{example}[Example~\ref{ex:pmultsss} continued]\label{ex:pmultssss}
  The set singleton $\{\rlbl{1}\}$ consisting of the dependency pair obtained from recursive addition rule 
  constitute trivially a forward closed set of dependency pairs. 
  Note that $\m{sep}(\{\rlbl{3})$ is given 
  \begin{align*}
    \rlbl{3a}\colon\ \ms(x) \mrk{\times} y & \to y \mrk{+} (x \times y) & 
    \rlbl{3b}\colon\ \ms(x) \mrk{\times} y & \to x \mrk{\times} y & 
  \end{align*}
  The following gives a sound inference
  $$
  \proc[]
    {\judge{\cp{\{\rlbl{1},\rlbl{3}\}}{\RSmult}{\RSmult}{\Tbs}}{f \cdot g}}
    {\judge{\cp{\{\rlbl{3}\}}{\RSmult}{\RSmult}{\Tbs}}{f}
    & \judge{\cp{\{\rlbl{1}\}}{\{\rlbl{3a},\rlbl{3b}\} \cup \RSmult}{\RSmult}{\Tbs}}{g}} \tpkt
  $$
  The generated sub-problem on the left is used to estimate applications of strict rule $\rlbl{3}$
  that occur only in the upper layer of derivation trees of $\PPmults$. 
  The sub-problem on the right is used to estimate rule $\rlbl{2}$
  of derivation trees from $t \in \Tbs$. 

  It is not difficult to find linear polynomial interpretations that verify that both sub-problems 
  have linear complexity. Overall the decomposition can thus prove the (asymptotically tight) bound 
  $\bigO(n^2)$ for the problem $\PPmults$, which in turn binds the complexity of $\PPmult$ 
  by Theorem~\ref{rc:proc:pe}, Theorem~\ref{rc:proc:rl} and Theorem~\ref{rc:proc:tuples}.
\end{example}

When the weak component of the considered DP problem contains dependency pairs, 
the situation gets slightly more involved. The following introduces dependency graph 
decomposition for this general case. 
Below the side condition $\Pre[\GS](\SSs_{\downarrow} \cup \WWs_{\downarrow}) \subseteq \SSs_{\uparrow}$
ensures that the bounding function $f$ accounts for the number of subtrees $\seq{T}$ in the lower layer, 
compare Figure~\ref{fig:dt:sep}.

\begin{theorem}[Dependency Graph Decomposition]\label{l:rc:dgcompose}
  Let $\PPs = \cp{\SSs \cup \SS}{\WWs \cup \WW}{\QQ}{\TTs}$ be a dependency problem, 
  and let $\GS$ denote the DG of $\PPs$.
  Let $\SSs_{\downarrow} \cup \SSs_{\uparrow } = \SSs$ and
  $\WWs_{\downarrow} \cup \WWs_{\uparrow } = \WWs$ be partitions
  such that $\SSs_{\downarrow } \cup \WWs_{\downarrow}$
  is closed under $\GS$-successors and $\Pre[\GS](\SSs_{\downarrow} \cup \WWs_{\downarrow}) \subseteq \SSs_{\uparrow}$.
  The following processor is sound.
  $$
  \proc[DG~decomp.]
  {\judge{\cp{\SSs \cup \SS}{\WWs \cup \WW}{\QQ}{\TTs}}{f \cdot g}}
  { \judge{\cp{\SSs_{\uparrow} \cup \SS}{\WWs_{\uparrow} \cup \WW}{\QQ}{\TTs}}{f}
    &
    \judge{\cp{\SSs_{\downarrow} \cup \SS}{\WWs_{\downarrow} \cup \m{sep}(\SSs_{\uparrow} \cup \WWs_{\uparrow}) \cup \WW}{\QQ}{\TTs}}{g}
  }
  \tpkt
  $$
\end{theorem}
\begin{proof}
  Let $\PP_{\uparrow} = \cp{\SSs_{\uparrow} \cup \SS}{\WWs_{\uparrow} \cup \WW}{\QQ}{\TT}$ 
  and $\PP_{\downarrow} = \cp{\SSs_{\downarrow} \cup \SS}{\WWs_{\downarrow} \cup \m{sep}(\SSs_{\uparrow} \cup \WWs_{\uparrow}) \cup \WW}{\QQ}{\TT}$
  where components are as given by the lemma. 
  Suppose $\cc[\PP_{\uparrow}](n) \in \bigO(f(n))$ and $\cc[\PP_{\downarrow}](n) \in \bigO(g(n))$. 
  According to Lemma~\ref{l:rc:dt} it is sufficient to show $\size[\SSs \cup \SS]{T} \in \bigO(f(n) \cdot g(n))$
  for any $\PP$-derivation tree $T$ of $t \in \TT$, where $t$ is of size up to $n$.
  Let $T_{\uparrow} = \cutAt{T}{\SSs_{\uparrow} \cup \WWs_{\uparrow} \cup \SS \cup \WW}$
  and denote by $\seq[m]{u}$ the \emph{trimmed inner nodes} in $T$ which are leafs in $T_{\uparrow}$.
  Let $T_i$ be the subtrees of $T$ rooted at $u_i$ ($i = 1,\dots,m$).
  By construction the root $u_i$ of $T_i$ is an $\SSs_{\downarrow} \cup \WWs_{\downarrow}$ node of $T_i$.
  As by assumption $\SSs_{\downarrow} \cup \WWs_{\downarrow}$ is closed under $\GS$-successors, 
  Lemma~\ref{l:rc:chain} yields that $T_i$ is an
  $\PP_\downarrow$ derivation tree.
  Consider the path $\pi$ from the root of $T$ to $u_i$ in $T$. By construction
  $\pi$ contains only $\SSs_{\uparrow} \cup \WWs_{\uparrow} \cup \SS  \cup \WW$ nodes.
  Using the dependency pairs in $\m{sep}(\SSs_{\uparrow} \cup \WWs_{\uparrow})$ and the rules from $\SS \cup \WW$ we can thus 
  extend $T_i$ to an $\PP_{\downarrow}$-derivation tree $T'_i$ of $t$.
  As $t \in \TT$ it follows that 
  \begin{align}
    \label{eq:dgcompose:down}
    \tag{\dag}
    \size[\SSs \cup \SS]{T_i} & = \size[\SSs_{\downarrow} \cup \SS]{T'_i} \in \bigO(g(n)) && (i = 1,\dots,m)
  \end{align}
  where the inclusion follows by Lemma~\ref{l:rc:dt} on the assumption $\cc[\PP_{\downarrow}](n) \in \bigO(g(n))$.

  Now consider $T_{\uparrow}$ which, by definition, is a $\PP_{\uparrow}$-derivation tree of $t$.
  Consequently 
  \begin{align}
    \label{eq:dgcompose:up}
    \tag{\ddag}
    \size[\SSs \cup \SS]{T_{\uparrow}} & = \size[\SSs_{\uparrow} \cup \SS]{T_{\uparrow}} \in \bigO(f(n)) && \phantom{(i = 1,\dots,m)}
  \end{align}
  follows using Lemma~\ref{l:rc:dt} on the assumption $\cc[\PP_{\uparrow}](n) \in \bigO(f(n))$.
  Recall that $T_\uparrow$ was obtained by trimming inner nodes $\seq[m]{u}$ in $T$.
  Suppose more than one subtree was trimmed, i.e., $m > 1$. 
  Hence $T$ is branching and contains at least a dependency pair. 
  Observe that trimmed inner nodes $u_i$ ($i = 1,\dots,m$) are $\SSs_{\downarrow} \cup \WWs_{\downarrow}$ nodes in $T$.
  Hence $m \leqslant \size[\SSs_{\downarrow} \cup \WWs_{\downarrow}]{T}$
  and by Lemma~\ref{l:rc:pre} we see 
  $\size[\SSs_{\downarrow} \cup \WWs_{\downarrow}]{T} 
  \leqslant \max\{1,\size[{\Pre[\GS](\SSs_{\downarrow} \cup \WWs_{\downarrow})}]{T} \cdot K \}$
  for $K$ the maximal arity of compound symbols in $\PP$. 
  Using the assumption that $\Pre[\GS](\SSs_{\downarrow} \cup \WWs_{\downarrow}) \subseteq \SSs_{\uparrow}$
  we conclude 
  $$
  \size[{\Pre[\GS](\SSs_{\downarrow} \cup \WWs_{\downarrow})}]{T} 
  \leqslant \size[\SSs_{\uparrow}]{T} 
  \leqslant \size[\SSs_{\uparrow}]{T_\uparrow} 
  \leqslant \size[\SSs_\uparrow \cup \SS]{T_\uparrow} \tpkt 
  $$
  Putting the equations together, we thus have
  $m \leqslant \max\{1, \size[\SSs_\uparrow \cup \SS]{T_\uparrow} \cdot K\}$. 
  
  Since $T$ was decomposed into prefix $T_{\uparrow}$ and subtrees $\seq[m]{T}$ we derive
  \begin{align*}
   \size[\SSs \cup \SS]{T} 
   & = \size[\SSs \cup \SS]{T_{\uparrow}} + \sum_{i=1}^m \size[\SSs \cup \SS]{T_i} \\
   & \leqslant \size[\SSs \cup \SS]{T_{\uparrow}} + \max\{1,\size[\SSs_{\uparrow} \cup \SS]{T_{\uparrow}} \cdot K\} \cdot \max\{\size[\SSs \cup \SS]{T_i}\mid i = 1,\dots,m\} \\
   % & \leqslant \size[\SSs \cup \SS]{T_{\uparrow}} + m \cdot \max\{\size[\SSs \cup \SS]{T_i}\mid i = 1,\dots,m\} \\
   & \leqslant \bigO(f(n)) + \bigO(f(n)) \cdot \bigO(g(n))
   && \text{by \eqref{eq:dgcompose:down}, \eqref{eq:dgcompose:up}} \\
   & = \bigO(f(n) \cdot g(n)) \tpkt
  \end{align*}
  As $T$ was an arbitrary $\PP$-derivation tree, the lemma follows by Lemma~\ref{l:rc:dt}.
\end{proof}

We remark that the inference given above in Example~\ref{ex:pmultssss} is an instance 
of dependency graph decomposition. 

\section{Conclusion}\label{s:rc:conclusion}
We have presented a combination framework for
polynomial complexity analysis of term rewrite systems.
The framework is general enough to reason about both runtime 
and derivational complexity, and to formulate a majority of 
the techniques available for proving polynomial complexity of rewrite systems. 
On the other hand, it is concrete enough to serve as a basis for a modular 
complexity analyser, as demonstrated by our automated complexity analyser \TCT\ 
which closely implements the discussed framework. 
Besides the combination framework we have introduced the notion 
of $\PP$-monotone complexity pair that unifies the different orders 
used for complexity analysis in the cited literature. 
Last but not least, we have presented the dependency graph decomposition 
processor. This processor is easy to implement, and greatly improves 
modularity.


\begin{thebibliography}{10}

\bibitem{AG:TCS:00}
T.~Arts and J.~Giesl.
\newblock {Termination of Term Rewriting using Dependency Pairs}.
\newblock {\em TCS}, 236(1--2):133--178, 2000.

\bibitem{A:ESSLLI:10}
M.~Avanzini.
\newblock {POP*} and {S}emantic {L}abeling using {SAT}.
\newblock In {\em Proc.\ of ESSLLI 2008/2009 Student Session}, volume 6211 of
  {\em LNCS}, pages 155--166. Springer, 2010.

\bibitem{AEM:TCS:12}
M.~Avanzini, N.~Eguchi, and G.~Moser.
\newblock {New Order-theoretic Characterisation of the Polytime Computable
  Functions}.
\newblock 2012.
\newblock Submitted to TCS.

\bibitem{AM:RTA:10}
M.~Avanzini and G.~Moser.
\newblock {Closing the Gap Between Runtime Complexity and Polytime
  Computability}.
\newblock In {\em Proc.\ of \nth{21} RTA}, volume~6 of {\em LIPIcs}, pages
  33--48, 2010.

\bibitem{AM:FLOPS:10}
M.~Avanzini and G.~Moser.
\newblock Complexity {A}nalysis by {G}raph {R}ewriting.
\newblock In {\em Proc.\ of \nth{10} FLOPS}, volume 6009 of {\em LNCS}, pages
  257--271. Springer, 2010.

\bibitem{AM:LMCS:12}
M.~Avanzini and G.~Moser.
\newblock {Polynomial Path Orders: A Maximal Model}.
\newblock 2012.
\newblock Submitted to LMCS.

\bibitem{BN98}
F.~Baader and T.~Nipkow.
\newblock {\em {Term Rewriting and All That}}.
\newblock Cambridge University Press, 1998.

\bibitem{BCMT:JFP:2001}
G.~Bonfante, A.~Cichon, J.-Y. Marion, and H.~Touzet.
\newblock Algorithms with polynomial interpretation termination proof.
\newblock {\em JFP}, 11(1):33--53, 2001.

\bibitem{HM:IC:05}
N.~Hirokawa and A.~Middeldorp.
\newblock {Automating the Dependency Pair Method}.
\newblock {\em IC}, 199(1--2):172--199, 2005.

\bibitem{HM:IJCAR:08}
N.~Hirokawa and G.~Moser.
\newblock Automated {C}omplexity {A}nalysis {B}ased on the {D}ependency {P}air
  {M}ethod.
\newblock In {\em Proc.\ of 4th IJCAR}, volume 5195 of {\em LNCS}, pages
  364--380, 2008.

\bibitem{HM:LPAR:08}
N.~Hirokawa and G.~Moser.
\newblock {Complexity, Graphs, and the Dependency Pair Method}.
\newblock In {\em Proc. of 15th LPAR}, pages 652--666, 2008.

\bibitem{HM:IC:12}
N.~Hirokawa and G.~Moser.
\newblock {Automated Complexity Analysis Based on the Dependency Pair Method}.
\newblock {\em IC}, 2012.
\newblock submitted.

\bibitem{HL:RTA:89}
D.~Hofbauer and C.~Lautemann.
\newblock {Termination Proofs and the Length of Derivations}.
\newblock In {\em Proc.\ of 3rd RTA}, volume 355 of {\em LNCS}, pages 167--177.
  Springer, 1989.

\bibitem{HW:RTA:06}
D.~Hofbauer and J.~Waldmann.
\newblock {Termination of String Rewriting with Matrix Interpretations}.
\newblock In {\em Proc.\ of \nth{17} RTA}, volume 4098 of {\em LNCS}, pages
  328--342. Springer, 2011.

\bibitem{L:SOFSEM:95}
S.~Lucas.
\newblock {Fundamentals of Context-Sensitive Rewriting}.
\newblock In {\em Proc.\ of \nth{22} SOFSEM}, LNCS, pages 405 -- 412. Springer,
  1995.
\newblock Creative-Commons-NC-ND licensed.

\bibitem{MMNWZ:CAI:11}
A.~Middeldorp, G.~Moser, F.~Neurauter, J.~Waldmann, and H.~Zankl.
\newblock {Joint Spectral Radius Theory for Automated Complexity Analysis of
  Rewrite Systems}.
\newblock In {\em Proc.\ of \nth{4} CAI}, volume 6742 of {\em LNCS}, pages
  1--20. Springer, 2011.

\bibitem{Moser:2009b}
G.~Moser.
\newblock Proof theory at work: Complexity analysis of term rewrite systems.
\newblock {\em CoRR}, abs/0907.5527, 2009.
\newblock Habilitation Thesis.

\bibitem{MSW:FSTTCS:2008}
G.~Moser, A.~Schnabl, and J.~Waldmann.
\newblock Complexity analysis of term rewriting based on matrix and context
  dependent interpretations.
\newblock In {\em Proc.\ of the 28th FSTTCS}, pages 304--315. LIPIcs, 2008.
\newblock Creative-Commons-NC-ND licensed.

\bibitem{NEG:CADE:11}
L.~Noschinski, F.~Emmes, and J.~Giesl.
\newblock {A Dependency Pair Framework for Innermost Complexity Analysis of
  Term Rewrite Systems}.
\newblock In {\em Proc.\ of 23rd CADE}, LNCS, pages 422--438. Springer, 2011.

\bibitem{AS:12}
A.~Schnabl.
\newblock {\em {Derivational Complexity Analysis Revisited}}.
\newblock PhD thesis, University of Innsbruck, 2012.

\bibitem{T:07}
R.~Thiemann.
\newblock {\em The {DP} Framework for Proving Termination of Term Rewriting}.
\newblock PhD thesis, University of Aachen, Department of Computer Science,
  2007.
\newblock available as Technical Report AIB-2007-17.

\bibitem{ZK:RTA:10}
H.~Zankl and M.~Korp.
\newblock {Modular Complexity Analysis via Relative Complexity}.
\newblock In {\em Proc.\ of \nth{21} RTA}, volume~6 of {\em LIPIcs}, pages
  385--400, 2010.

\bibitem{Z:RTA:97}
H.~Zantema.
\newblock {Termination of Context-Sensitive Rewriting}.
\newblock In {\em Proc.\ of \nth{8} RTA}, volume 1232 of {\em LNCS}, pages
  172--186. Springer, 1997.

\end{thebibliography}
\end{document}